\documentclass[10pt,english,reprint, aps, pra, superscriptaddress, nofootinbib]{revtex4-1}
\usepackage[T1]{fontenc}
\usepackage[latin9]{inputenc}
\usepackage{geometry}
\usepackage{color}
\usepackage{babel}
\usepackage{float}
\usepackage{amsmath}
\usepackage{amssymb}
\usepackage{graphicx}
\usepackage[unicode=true,pdfusetitle,
 bookmarks=true,bookmarksnumbered=false,bookmarksopen=false,
 breaklinks=false,pdfborder={0 0 0},pdfborderstyle={},backref=false,colorlinks=true]
 {hyperref}
\hypersetup{
 colorlinks=true, allcolors=darkblue}

\makeatletter
\usepackage{xcolor}
\definecolor{darkblue}{rgb}{0.0, 0.0, 0.55}

\makeatother

\begin{document}
\title{Generation of Perfectly Achromatic Optical Vortices Using a Compensated
Tandem Twisted Nematic Cell}
\author{Dmytro O. Plutenko}
\email{dmytro.plutenko@gmail.com}

\address{Institute of Physics of National Academy of Sciences of Ukraine, Prospekt
Nauky 46, 03680 Kyiv, Ukraine}
\author{Mikhail V. Vasnetsov}
\email{vasnet@hotmail.com}

\address{Institute of Physics of National Academy of Sciences of Ukraine, Prospekt
Nauky 46, 03680 Kyiv, Ukraine}
\begin{abstract}
The generation of ``white'' optical vortices is currently constrained
by intrinsic trade-offs between spectral bandwidth, conversion efficiency,
and temporal pulse integrity in conventional diffractive and geometric-phase
approaches. In this work, we theoretically investigate a compensated
tandem crossed twisted nematic (TN) liquid crystal architecture that
overcomes these fundamental limitations. By developing a rigorous
Jones matrix model and defining specific figures of merit for chromatic
fidelity, we analyze the impact of manufacturing imperfections and
non-adiabatic waveguiding (deviations from the Mauguin regime) on
the device performance. We propose and evaluate three distinct compensation
strategies, ranging from optimized passive designs for specific manufacturing
tolerances to a robust active compensation scheme utilizing a tunable
retarder. Our analysis demonstrates that the active approach effectively
nullifies parasitic amplitude modulation, enabling the generation
of perfectly achromatic vortices with high phase purity across an
arbitrary bandwidth. This establishes the compensated tandem TN cell
as a superior and versatile platform for high-fidelity white-light
singular optics.
\end{abstract}
\maketitle

\section{Introduction}

Optical vortices are a fascinating class of structured light beams
characterized by a helical phase front, described by the term $\exp\left(il\phi\right)$,
where $l$ is an integer known as the topological charge and $\phi$
is the azimuthal angle. This helical phase results in a phase singularity
on the beam axis, leading to a characteristic donut-shaped intensity
profile. A landmark discovery by Allen et al. in 1992 revealed that
such beams carry a well-defined orbital angular momentum (OAM) of
$l\hbar$ per photon, distinct from the spin angular momentum associated
with polarization \citep{PhysRevA.45.8185,PhysRevA.56.4064,BEIJERSBERGEN1993123,PADGETT199536}.
This property enables the transfer of torque to microscopic particles,
fueling a vast range of applications, including advanced optical tweezers
for manipulating and rotating particles \citep{ROTATION_doi:10.1021/acsphotonics.5c00137,ROTATION_manipulation_Grier,PhysRevA.Kiselev.Plutenko.2016,PhysRevLett.75.826,PhysRevA.54.1593,Padgett2011},
and broadband super-resolution imaging where achromatic depletion
beams enable multi-color stimulated emission depletion microscopy
\citep{Wildanger:08,Yan:15}, increasing channel capacity in optical
communications through mode-division multiplexing \citep{multiplexing_Willner:15,Wang2012},
and pushing the boundaries of quantum information science with a high-dimensional
state space \citep{quantum_protocols}.

The utility of these applications is significantly enhanced when they
can be performed with broadband light, which has spurred intense interest
in \textquotedbl white light\textquotedbl{} or achromatic optical
vortices \citep{Generation_OV_Yao:11,Denisenko:09,Mariyenko:05,Leach:06,Swartzlander06}.
A truly achromatic vortex generator must ideally preserve the key
topological properties across a broad spectral range. This implies
not only topological achromaticity (a constant integer charge $l$)
but also device achromaticity. For a device, this means providing
wavelength-independent phase modulation and, crucially, maintaining
a uniform amplitude response without introducing chromatic power losses.
While the diffractive nature of light means that a beam's size will
inevitably scale with wavelength, the generating element itself should
ideally offer a spatially and spectrally uniform transformation.

However, generating high-quality white vortices remains a significant
challenge, as most conventional generation methods are inherently
chromatic. Techniques based on diffractive optical elements, such
as spiral phase plates and forked gratings, introduce a phase shift
proportional to the optical path difference, which is fundamentally
dependent on wavelength \citep{Generation_OV_Yao:11,BASISTIY1993422}.
While ubiquitous, standard liquid crystal spatial light modulators
(SLMs) operate on the same principle and thus suffer from the same
chromatic limitations \citep{SLM_review_oes-2023-0026,Calero13,Leach:06}.
Alternative approaches using dielectric metasurfaces have demonstrated
broadband capabilities \citep{Khorasaninejad2017,Wang2017,Qian:23},
yet they often face challenges related to fabrication complexity and
static response profiles \citep{Capasso+2018+953+957}.

The most promising route to tunable achromaticity has been through
devices that impart a geometric (Pancharatnam-Berry) phase \citep{Q-platePhysRevLett.96.163905}.
Q-plates, typically fabricated from spatially patterned liquid crystals,
are the leading candidates in this category \citep{Slussarenko:11,Rubano:19,Karimi09,10.1063/1.3527083}.
They function as spatially variant half-wave plates that convert the
spin of circularly polarized light into OAM. While the geometric phase
itself is achromatic, the underlying half-wave plate condition is
not, constraining broadband performance. Numerous strategies have
been proposed to create achromatic half-wave plates, for instance
by stacking multiple retarders \citep{Komanduri:13,Pancharatnam1955}
or using Bragg reflection in helical liquid crystal structures \citep{Breg_berry_mirror_1_Rafayelyan:16,Breg_berry_mirror_PhysRevA.96.043862},
but these methods are often complex and provide only quasi-achromatic
performance over a limited bandwidth. A critical drawback is the induction
of wavelength-dependent group delay dispersion, leading to nonlinear
temporal pulse shaping that is difficult to mitigate with standard
dispersion compensation techniques. A particularly elegant approach
involves using a standard, chromatic q-plate followed by a polarization
filtering system \citep{Q_plate_filteringGecevicius:18}. This method
successfully achieves a perfectly achromatic phase response, but at
a significant cost: it introduces a strong, non-uniform amplitude
modulation, as unwanted polarization components at off-design wavelengths
are filtered out, leading to chromatic power loss.

In this work, we introduce and analyze a fundamentally new approach
to this problem. Our method is based on a device architecture, previously
known only for its application as a broadband polarization rotator,
which consists of a tandem of two twisted nematic (TN) liquid crystal
cells \citep{achromatic_rotator}. To the best of our knowledge, the
potential of this tandem cell architecture as a high-performance vortex
generator has not been previously explored. We demonstrate that, with
a proposed active compensation scheme, this system can overcome the
critical limitations of prior art. It can be engineered to provide
not only a perfectly achromatic phase response but also a fully achromatic
amplitude transmission, thus eliminating the parasitic intensity modulation
that has constrained previous geometric phase optics. Herein, we develop
a comprehensive theoretical model for this device, analyze the sources
of non-idealities, and present three distinct compensation strategies
that establish the tandem TN cell as a superior and highly versatile
platform for generating high-fidelity white optical vortices.

\section{Theoretical Model}

\subsection{Light Propagation in a Single Twisted Nematic Cell}

Let us consider a single liquid crystal cell with a planar alignment
of the director. We will choose the Cartesian coordinate system such
that the $z$-axis is orthogonal to the two parallel substrates of
the cell. The standard-basis vectors of the coordinate system we denote
as $\left\{ \hat{\mathbf{e}}_{x},\hat{\mathbf{e}}_{y},\hat{\mathbf{e}}_{z}\right\} $.
The orientation of the director of the liquid crystal can be defined
by a unit vector which we denote as $\hat{\mathbf{e}}$. We assume
that the alignment of the director is planar, therefore the vector
$\hat{\mathbf{e}}$ can be defined by a twist angle $\phi$, which
is an angle between axis $x$ and director orientation. 
\begin{equation}
\hat{\mathbf{e}}=\hat{\mathbf{e}}_{x}\cos\phi+\hat{\mathbf{e}}_{y}\sin\phi\label{eq:e_fi}
\end{equation}

We model the propagation of light along the $z$-axis through the
liquid crystal by treating it as a birefringent medium. The model
is built upon two principal refractive indices: the ordinary index,
$n_{o}$, and the extraordinary index, $n_{e}$, which are assumed
to be constant throughout the medium.

The local orientation of the director determines the optical axes
for a light ray. Specifically, the extraordinary wave is polarized
parallel to the director, while the ordinary wave is polarized orthogonally
to it.

This framework is well-suited for configurations where the director
varies slowly. However, it can also effectively model systems with
sharp interfaces by considering them as a stack of discrete, uniform
layers. Thus, for a single homogeneous layer of thickness $h$, the
light propagation can be described by the following equation

\begin{flalign}
\begin{split}\mathbf{E}\left(z+h\right) & =\left(\mathbf{E}\left(z\right)\cdot\hat{\mathbf{e}}\right)\hat{\mathbf{e}}\exp\left(in_{e}kh\right)+\\
+ & \left[\mathbf{E}\left(z\right)-\left(\mathbf{E}\left(z\right)\cdot\hat{\mathbf{e}}\right)\hat{\mathbf{e}}\right]\exp\left(in_{o}kh\right)
\end{split}
\label{eq:1_Ezh}
\end{flalign}
where $\mathbf{E}$ is a vector of electric field, $k$ is a wave
number in vacuum. For the subsequent analysis, it is convenient to
use the Jones formalism. Within this framework, transmission through
the layer $h$ is described by the Jones matrix $T^{E}$$\left(h\right)$.
\begin{equation}
\mathbf{E}\left(z+h\right)=T_{h}^{E}\left(h\right)\mathbf{E}\left(z\right)\label{eq:E=00003DTE}
\end{equation}
\begin{equation}
\begin{aligned}T^{E}\left(h\right)= & \exp\left(i\bar{k}h\right)\left[\begin{pmatrix}1 & 0\\
0 & 1
\end{pmatrix}\cos\left(\gamma h\right)+\right.\\
+ & \left.i\begin{pmatrix}\cos\left(2\phi\right) & \sin\left(2\phi\right)\\
\sin\left(2\phi\right) & -\cos\left(2\phi\right)
\end{pmatrix}\sin\left(\gamma h\right)\right]
\end{aligned}
\label{eq:Th-1}
\end{equation}
\begin{equation}
\bar{k}=\frac{n_{e}+n_{0}}{2}k\label{eq:kmean}
\end{equation}
\begin{equation}
\gamma=\frac{\Delta n}{2}k=\frac{n_{e}-n_{0}}{2}k\label{eq:gamma}
\end{equation}
where $T^{E}$ is a Jones transmission matrix, $\bar{k}$ is a mean
wave number in the liquid crystal, $\gamma$ is an anisotropy parameter. 

For the matrix formulation in Eq.~(\ref{eq:E=00003DTE}), we express
the vector $\mathbf{E}$ as a column vector of its components in the
Cartesian basis.

To better analyze the evolution of the polarization state, it is convenient
to switch from the linear (Cartesian) basis to the circular polarization
basis. We introduce a new Jones vector, $\boldsymbol{\mathcal{E}}$,
which is related to the original vector $\mathbf{E}$ through a basis
transformation matrix $U$. Concurrently, we separate out the common
phase factor that corresponds to the average optical path through
the liquid crystal. The resulting transformation is given by:

\begin{equation}
\boldsymbol{\mathcal{E}}\left(z\right)=\exp\left(-i\bar{k}z\right)U\mathbf{E}\left(z\right)\label{eq:=000415=000415-1}
\end{equation}
\begin{equation}
U=\frac{1}{\sqrt{2}}\begin{pmatrix}1 & i\\
1 & -i
\end{pmatrix}\label{eq:U}
\end{equation}
With the Jones vector $\boldsymbol{\mathcal{E}}$ now expressed in
the circular basis, its components $\mathcal{E}_{1}$ and $\mathcal{E}_{2}$
correspond to the complex amplitudes of the right-hand circular (RHC)
and left-hand circular (LHC) polarizations, respectively. The transmission
through the system is described by the Jones matrix in this basis,
$T^{\mathcal{E}}$, given by:

\begin{equation}
T^{\mathcal{E}}=\exp\left(-i\bar{k}z\right)UT^{E}U^{-1}\label{eq:TE}
\end{equation}
\begin{equation}
\begin{aligned}T^{\mathcal{E}}\left(h\right)= & \begin{pmatrix}1 & 0\\
0 & 1
\end{pmatrix}\cos\left(\gamma h\right)+\\
+ & i\begin{pmatrix}0 & \exp\left(i2\phi\right)\\
\exp\left(-i2\phi\right) & 0
\end{pmatrix}\sin\left(\gamma h\right)
\end{aligned}
\label{eq:TE2}
\end{equation}
The matrix $T^{\mathcal{E}}$ describes light propagation through
a uniform layer. To generalize this for a non-uniform liquid crystal,
where the director orientation varies continuously along the $z$-axis,
we now consider an infinitely thin layer of thickness $dz$. By taking
the limit of the transfer matrix $T^{\mathcal{E}}\left(h\right)$
as $h\rightarrow0$, we can obtain a differential equation that governs
the continuous evolution of the light's polarization state.
\begin{equation}
\frac{d}{dz}\boldsymbol{\mathcal{E}}\left(z\right)=i\gamma\begin{pmatrix}0 & \exp\left(i2\phi\right)\\
\exp\left(-i2\phi\right) & 0
\end{pmatrix}\boldsymbol{\mathcal{E}}\left(z\right)\label{eq:DifE}
\end{equation}

While in the general case, Eq.~(\ref{eq:DifE}) does not have a simple
analytical solution due to the non-commuting nature of the corresponding
matrix in the equation, a closed-form solution can be found for the
important case of a linear director twist, $\phi\left(z\right)=\phi_{0}+qz$.
This configuration corresponds to a uniformly twisted nematic or cholesteric
structure.

The standard procedure for this case involves reducing the system
of two first-order coupled differential equations to a single equivalent
second-order linear homogeneous equation with constant coefficients.
Since this is a well-established method for solving this class of
problems, we will omit the detailed derivation and present the final
analytical expression for the system's Jones matrix $T_{\text{TN}}^{\mathcal{E}}\left(h\right)$.

\begin{widetext}

\begin{equation}
T_{\text{TN}}^{\mathcal{E}}\left(h\right)=\begin{pmatrix}\exp\left(i\phi_{d}\right)\left[\cos\left(\gamma^{'}h\right)-i\frac{q}{\gamma^{'}}\sin\left(\gamma^{'}h\right)\right] & i\frac{\gamma}{\gamma^{'}}\exp\left(i2\bar{\phi}\right)\sin\left(\gamma^{'}h\right)\\
i\frac{\gamma}{\gamma^{'}}\exp\left(-i2\bar{\phi}\right)\sin\left(\gamma^{'}h\right) & \exp\left(-i\phi_{d}\right)\left[\cos\left(\gamma^{'}h\right)+i\frac{q}{\gamma^{'}}\sin\left(\gamma^{'}h\right)\right]
\end{pmatrix}\label{eq:T_TN}
\end{equation}
\end{widetext}where $h$ is a layer thickness, $q$ is a twist wave
number ($q=h^{-1}\left[\phi\left(z+h\right)-\phi\left(z\right)\right]$)
, $\bar{\phi}$ is a mean value of the director's orientation angle
$\bar{\phi}=\frac{1}{2}\left[\phi\left(z\right)+\phi\left(z+h\right)\right]=\phi\left(z\right)+\frac{1}{2}qh$,
$\phi_{d}$ is full angle of rotation of the director in the layer
\begin{equation}
\phi_{d}=\phi\left(z+h\right)-\phi\left(z\right)=qh\label{eq:fd}
\end{equation}
\begin{equation}
\gamma^{'}=\sqrt{\gamma^{2}+q^{2}}\label{eq:gamma1}
\end{equation}

\subsection{The Ideal Tandem Cell as an Achromatic Rotator}

The tandem twisted nematic (TN) liquid crystal cell is a known device,
previously investigated for its properties as a broadband polarization
rotator \citep{achromatic_rotator}. In this work, we explore a novel
application for this system: its potential for generating optical
vortices, particularly with broadband (white) light. Our goal is to
analyze the performance and limitations of this tandem rotator when
used to create the spatially variant polarization rotation required
for vortex formation, thus examining its viability in a role often
fulfilled by elements like q-plates.

\begin{figure}[H]
\includegraphics[width=0.9\columnwidth]{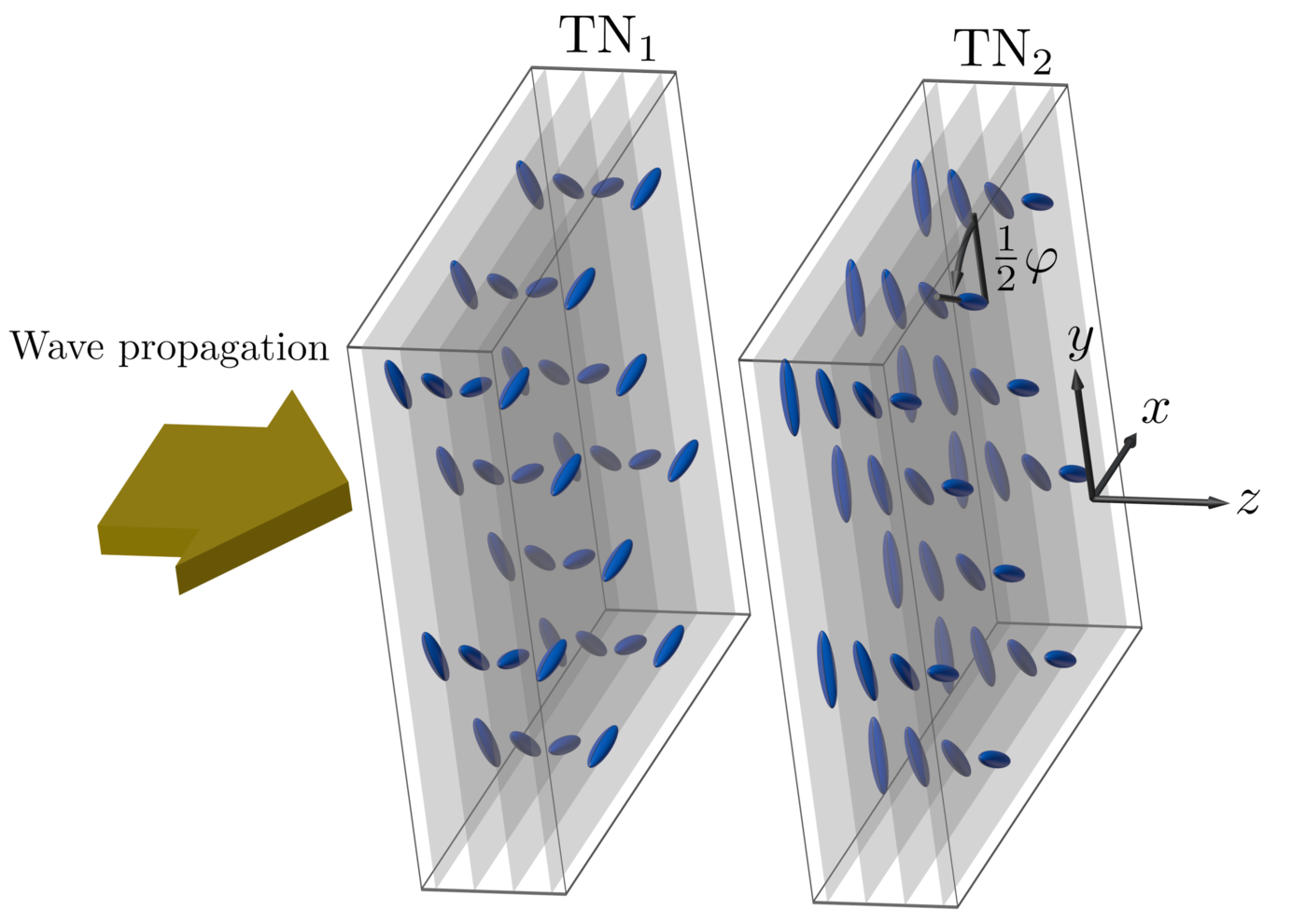}\caption{\label{fig:2}Configuration of the tandem TN cell system.}
Two TN cells, $\text{TN}_{1}$ and $\text{TN}_{2}$, are cascaded.
The director at the output of $\text{TN}_{1}$ is aligned with the
$x$-axis, while the director at the input of $\text{TN}_{2}$ is
aligned with the $y$-axis, creating an orthogonal interface.
\end{figure}

To establish the fundamental principle, we first analyze an idealized
model. We consider a system composed of two identical TN cells, whose
individual behavior is described by the Jones matrix $T_{\text{TN}}^{\mathcal{E}}$.
The cells are assumed to have identical material properties, thickness
$h$, and the twist angle $\frac{\varphi}{2}$ (the total twist angle
for two cells together is $\varphi$). They are cascaded such that
the director at the output of the first cell is orthogonal to the
director at the input of the second (as illustrated in Fig.~\ref{fig:2}).

The key to this analysis is the assumption that the system operates
perfectly within the Mauguin (wave-guiding) regime. In this ideal
limit, the device functions as a pure polarization rotator. The mechanism
for creating an optical vortex relies on making the angle of this
rotation spatially dependent across the beam's profile. Such a spatially
varying rotation leads to the phase modulation of the the circular
polarized incident light, which transforms a planar wavefront into
the helical wavefront characteristic of an optical vortex.

We begin by analyzing this idealized case to formally demonstrate
how the tandem cell's rotational properties can be harnessed for this
purpose. We will derive the Jones matrix for the ideal tandem rotator,
which will serve as a baseline model. This analysis is a necessary
first step before investigating the significant practical challenges
and non-ideal effects, such as deviations from the Mauguin regime
and manufacturing imperfections, which are the primary focus of our
subsequent research.

The Mauguin (or wave-guiding) regime is defined by the condition $q\ll\gamma^{'}$,
where the twist wave number is much smaller than the liquid crystal's
anisotropy parameter. Physically, this implies that the director twist
is very gradual, allowing the polarization of light to adiabatically
follow its orientation. Under this approximation, the term $\gamma^{'}$
in Eq.~(\ref{eq:gamma1}) simplifies to $\gamma^{'}\approx\gamma$,
which significantly simplifies the Jones matrix for a single TN cell.

As described, our tandem cell consists of two identical TN cells,
each with a twist of $\frac{\varphi}{2}$, yielding a total twist
of $\varphi$. The resulting Jones matrix for the entire system (two
crossed TN LC cells), $T_{\text{CTN}}^{\mathcal{E}}$, is obtained
by multiplying the matrices of the individual cells, taking into account
that the director at the input of the second cell is orthogonal to
the director at the output of the first. This yields the following
expression: 
\begin{equation}
T_{\text{CTN}}^{\mathcal{E}}=\begin{pmatrix}\exp\left(i\varphi\right) & 0\\
0 & \exp\left(-i\varphi\right)
\end{pmatrix}\label{eq:TE_CTN}
\end{equation}
A key insight provided by this resulting matrix is its effect on circularly
polarized light. It functions by applying a phase shift of $\varphi$
to the right-hand circular (RHC) polarization component and $-\varphi$
to the left-hand circular (LHC) component. Crucially, the transformation
depends only on the total twist angle $\varphi$ of the director profile
and is independent of other physical parameters like cell thickness
or initial orientation.

This mechanism is fundamentally different from that of a $q$-plate.
While a $q$-plate flips the helicity of circular polarizations (RHC
becomes LHC and vice-versa), this tandem cell preserves it: RHC light
remains RHC, and LHC remains LHC, with the device solely modulating
their respective phases.

A remarkable feature of this result is that the imparted phase modulation
$\pm\varphi$ is inherently independent of wavelength. Although a
wavelength-dependent phase factor was separated in our derivation
Eq.~(\ref{eq:=000415=000415-1}), it represents only the average
optical path through the cell and does not influence the differential
phase modulation responsible for the vortex formation. This means
that, in its ideal form, the tandem cell acts as a perfectly achromatic
device. It can transform a circularly polarized plane wave into an
optical vortex across a broad spectrum (i.e., a ``white'' vortex),
with the sole condition that the Mauguin regime is maintained for
all relevant wavelengths.

\section{\label{sec:Analysis-of-a-TNC}Analysis of a Non-Ideal Tandem Cell}

The reliance on the strict Mauguin regime, as discussed in the ideal
model, presents the central challenge for practical implementation.
In any real-world device, achieving this regime perfectly across a
wide spectral range is practically impossible. Thus, we must investigate
how the device's performance is affected by non-idealities. Specifically,
we will analyze two primary sources of error: 1) the effect of imperfect
wave-guiding due to deviations from the strict Mauguin condition,
and 2) manufacturing tolerances, particularly the mismatch in the
thicknesses of the two cells.

It is crucial to clarify the scope of our analysis here. We are investigating
how these physical device imperfections affect the generation of the
phase modulation profile itself. We are not analyzing the subsequent,
well-documented problem of how an imperfectly shaped phase profile
(for example, a profile with a phase step due to insufficient modulation)
impacts the resulting vortex, which is known to decompose into a superposition
of vortex states. Our focus is squarely on the physical origins of
phase and amplitude distortions arising from the non-ideal behavior
of this specific tandem cell architecture.

\subsection{Perturbation Model and Small Parameters}

To construct the model, we consider a tandem system composed of two
TN cells made from the same liquid crystal material. The cells have
thicknesses of $h_{1}$ and $h_{2}$, respectively, and each imparts
a director twist of $\varphi/2$, resulting in a total twist of $\varphi$
for the system. The relative orientation of the cells is fixed such
that the director at the output of the first cell is aligned along
the $x$-axis, while the director at the input of the second cell
is aligned along the $y$-axis.

While an exact analytical expression for the total Jones matrix of
this system can be derived, it is excessively cumbersome and offers
little direct physical insight. A more powerful approach is to analyze
the system's behavior by expanding the solution in terms of small,
dimensionless parameters that characterize the primary sources of
non-ideality.

First, to quantify the deviation from the ideal Mauguin regime, we
introduce the mean parameter $\alpha$ using the effective wavenumber
$\gamma'=\sqrt{\gamma^{2}+q^{2}}$: 
\begin{equation}
\alpha=\frac{1}{2}\left(\frac{q_{1}}{\gamma_{1}'}+\frac{q_{2}}{\gamma_{2}'}\right)
\end{equation}
The second, $\beta=\left(h_{1}-h_{2}\right)/\left(h_{2}+h_{1}\right)$,
characterizes the relative mismatch in the cell thicknesses.

While an exact analytical expression for the total Jones matrix can
be derived, it is cumbersome. To facilitate a rigorous analysis capturing
both phase and amplitude effects, we perform a perturbative expansion
accurate up to the second order in these small parameters, as detailed
in Appendix \ref{subsec:Supplementary_2}. For the sake of clarity
in the main text, we will present the resulting expressions in their
compact forms, highlighting the dominant physical terms.\begin{widetext}

\begin{equation}
T_{\text{CTN}}^{\mathcal{E}}=\begin{pmatrix}\exp\left(i\varphi\right)\left[\cos\left(\Phi_{d}\right)-i\alpha\sin\left(\Phi_{s}\right)\right] & i\sin\left(\Phi_{d}\right)+2\alpha\sin\left(\gamma_{2}^{'}h_{2}\right)\sin\left(\gamma_{1}^{'}h_{1}\right)\\
i\sin\left(\Phi_{d}\right)-2\alpha\sin\left(\gamma_{2}^{'}h_{2}\right)\sin\left(\gamma_{1}^{'}h_{1}\right) & \exp\left(-i\varphi\right)\left[\cos\left(\Phi_{d}\right)+i\alpha\sin\left(\Phi_{s}\right)\right]
\end{pmatrix}\label{eq:TE_CTN2}
\end{equation}
\end{widetext}Here, the arguments $\Phi_{s}$ and $\Phi_{d}$ compactly
describe the phase properties of the tandem system. $\Phi_{s}$ represents
the cumulative phase retardation, which is the sum of the phase differences
accumulated between the wave-guided eigenmodes in each of the two
cells, given by

\begin{equation}
\Phi_{s}=\gamma_{1}^{'}h_{1}+\gamma_{2}^{'}h_{2}\label{eq:FS}
\end{equation}
Conversely, $\Phi_{d}$ represents the retardation mismatch between
the two cells and is defined as
\begin{equation}
\Phi_{d}=\gamma_{1}^{'}h_{1}-\gamma_{2}^{'}h_{2}\label{eq:FD}
\end{equation}
This parameter quantifies the degree of phase imbalance in the system. 

The small dimensionless parameters, $\ensuremath{\alpha}$ and $\ensuremath{\beta}$,
introduced previously in this section, are now intuitively linked
to these phase arguments. The thickness mismatch parameter, $\ensuremath{\beta}$,
is directly proportional to the phase imbalance: $\ensuremath{\beta\approx\Phi_{d}/\Phi_{s}}$.
Similarly, the non-ideal wave-guiding effect, $\ensuremath{\alpha}$,
can be related to the total director twist $\ensuremath{\varphi}$
and the cumulative retardation $\ensuremath{\Phi_{s}}$ as $\ensuremath{\alpha\approx\varphi/\Phi_{s}}$.
This intuitive relation demonstrates that non-ideal effects are suppressed
when the cumulative retardation is significantly larger than the total
director twist angle, thereby explicitly highlighting the Mauguin
regime condition.

\subsection{\label{subsec:First-Order-Solution-and}Transmission Analysis and
Filtering of Parasitic Components}

Analyzing the effect of the non-idealities reveals a key difference
from the ideal case. When circularly polarized light (e.g., RHC) passes
through the system, a parasitic, orthogonally polarized (LHC) component
emerges in addition to the primary RHC wave. This unwanted component
lacks the desired vortex-forming phase modulation and would otherwise
degrade the quality of the output beam.

Fortunately, the parasitic component can be effectively filtered out
using a standard polarimetric setup similar to that proposed in \citep{Q_plate_filteringGecevicius:18}.
The optical configuration matches the input and output stages depicted
in Fig. \ref{fig:3} (noting that for the uncompensated analysis in
this section, the compensating phase plate (CPP) is omitted). The
optical train consists of an input circular polarization generator
(a linear polarizer followed by an achromatic quarter-wave plate)
and an output circular polarization analyzer (a second achromatic
quarter-wave plate followed by a linear analyzer).

A key distinction dictates the specific configuration of the output
stage. The $q$-plate described in \citep{Q_plate_filteringGecevicius:18}
acts as a half-wave plate, inverting the polarization handedness (e.g.,
transforming right-handed circular (RHC) to left-handed circular (LHC)).
Consequently, their output stage is configured to transmit the orthogonal
polarization (LHC). In contrast, our tandem TN cell operates as a
polarization rotator, preserving the incident handedness (RHC remains
RHC, with only a phase modulation). Therefore, the output filtering
stage must be configured to transmit the same circular polarization
state as generated at the input. This implies that the relative orientation
of the final Analyzer with respect to the second quarter-wave plate
(QWP) must be set to pass the incident helicity, whereas the $q$-plate
scheme requires a configuration that isolates the orthogonal (opposite)
helicity.

The complex transmission coefficient, $T_{\text{CTN}}$, for the initial,
linearly polarized wave passing through this entire optical system
is derived in Appendix \ref{subsec:Supplementary_2} Eqs.~(\ref{eq:T11-2})--(\ref{eq:T22-2}).
Retaining the primary terms governing the device physics, the transmission
can be effectively approximated as: 
\begin{equation}
T_{\text{CTN}}=\exp\left(i\varphi\right)\left[\cos\left(\Phi_{d}\right)-i\alpha\sin\left(\Phi_{s}\right)\right]\label{eq:T_CTN}
\end{equation}
In the limit where $\left|\cos\left(\Phi_{d}\right)\right|\gg\alpha$,
the transmission coefficient can be approximated in a form that explicitly
separates its amplitude and phase components:
\begin{equation}
T_{\text{CTN}}\approx\cos\left(\Phi_{d}\right)\exp\left(i\varphi\left[1-\frac{\sin\left(\Phi_{s}\right)}{\Phi_{s}\cos\left(\Phi_{d}\right)}\right]\right)\label{eq:T_CTN2}
\end{equation}

The derived expression for the transmission coefficient Eq.~(\ref{eq:T_CTN2})
provides a basis for analyzing the device performance. While the first-order
approximation suggests that amplitude modulation depends solely on
the retardation mismatch $\Phi_{d}$, a rigorous second-order analysis
(derived in Sec. \ref{subsec:Supplementary_2}) reveals the complete
dependence. Retaining terms of order $\alpha^{2}$, the transmission
magnitude is given by:

\begin{equation}
\begin{aligned}\left|T_{\text{CTN}}\right|=\cos\left(\Phi_{d}\right)+ & \frac{1}{2}\alpha^{2}\frac{\sin^{2}\left(\Phi_{s}\right)}{\cos\left(\Phi_{d}\right)}-\\
- & 2\alpha^{2}\sin\left(\gamma_{1}^{'}h_{1}\right)\sin\left(\gamma_{2}^{'}h_{2}\right)
\end{aligned}
\label{eq:mod_T_exact-1}
\end{equation}

This expression highlights that the amplitude response is governed
by two factors: the retardation mismatch $\Phi_{d}$ (arising from
manufacturing tolerances) and the finite Mauguin parameter $\alpha$
(arising from non-adiabatic waveguiding).

In the ideal case of perfectly matched cells ($\beta=0$, thus $\Phi_{d}=0$)
and infinite thickness ($\alpha\to0$), the amplitude term becomes
unity. However, in a real device, the system introduces wavelength-dependent
losses. For the regime of interest --- thick cells designed to minimize
phase distortion --- the parameter $\alpha$ becomes small ($\alpha\propto1/h$),
whereas the sensitivity to retardation mismatch $\Phi_{d}$ increases
($\Phi_{d}\approx\beta\Phi_{s}\propto h$). Consequently, the $\cos\left(\Phi_{d}\right)$
term typically dominates the power loss mechanism, while the $\alpha^{2}$
correction provides a minor contribution.

The phase response of the system consists of two parts: the desired
linear phase shift proportional to the total director twist angle
$\varphi$, and an unwanted perturbation term. In the ideal Mauguin
regime ($\alpha=0$), this perturbation vanishes, and the imparted
phase is exactly equal to $\varphi$. For a non-zero $\alpha$, a
phase distortion appears. The amplitude of this distortion is directly
proportional to $\alpha$, and its value oscillates as a function
of wavelength with a periodicity determined by the sum phase retardation,
$\Phi_{s}$.

In the limit of small $\alpha$ and $\beta$, the relationship $\beta\approx\Phi_{d}/\Phi_{s}$
holds. If $\Phi_{s}$ is large, the phase perturbation becomes a rapidly
oscillating function of wavelength. To suppress this phase distortion,
one must reduce $\alpha$, which can be achieved by increasing the
cell thickness.

However, this presents a significant practical trade-off. While increasing
the cell thickness improves the phase fidelity, it simultaneously
makes it more challenging to maintain a small absolute thickness difference
($h_{1}-h_{2}$). A larger absolute difference leads to a larger $\Phi_{d}$,
which in turn increases the chromatic power losses from the amplitude
modulation term. Thus, a manufacturer would face a compromise: improving
the phase response at the cost of worsening the amplitude response.

\subsection{\label{subsec:Figures-of-Merit}Figures of Merit for Non-Achromaticity}

To quantitatively evaluate the device's worst-case performance, we
define specific figures of merit (FoM) for achromatic fidelity. These
metrics are expressed as the maximum deviation across the operational
spectral bandwidth ($\ensuremath{\Delta\lambda}$) and depend on fundamental
device and manufacturing parameters.

\paragraph{Amplitude Non-Achromaticity ($\ensuremath{\mathcal{A}}$)}

The amplitude non-achromaticity ($\mathcal{A}$) serves as the spectral
envelope for the power loss. It is defined as the local upper bound
for the fractional reduction in amplitude transmission at a given
wavelength $\lambda$:

\begin{equation}
\ensuremath{\mathcal{A}}=\max_{\text{local}}\left(1-\left|T\left(\lambda\right)\right|\right)\label{eq:A_achromaticity}
\end{equation}
where $T$ is a transmission coefficient of the system. This metric
estimates the worst-case loss magnitude for the current optical parameters
given the manufacturing tolerances, without averaging over the spectral
bandwidth.

\paragraph{Phase Non-Achromaticity}

Similarly, the phase non-achromaticity quantifies the envelope of
the chromatic error in the generated phase shift $\delta\varphi=\arg\left(T\right)-\varphi$.
To characterize the modulator's fidelity, we define the maximum phase
distortion magnitude at a given wavelength by the following linear
relationship: 
\begin{equation}
\max_{\text{local}}\left|\delta\varphi\left(\lambda,\varphi\right)\right|\approx\mathcal{P}_{0}+\left|\varphi\right|\mathcal{P}_{1}\label{eq:P_achromaticity}
\end{equation}

where $\mathcal{P}_{0}$ and $\mathcal{P}_{1}$ are the spectral envelopes
for the constant and angular phase distonstant Phase Distortions,
respectively.

\subsection{Non-Achromaticity of the Non-Ideal Tandem Cell}

The complex transmission coefficient Eqs.~(\ref{eq:T_CTN})--(\ref{eq:T_CTN2})
describes the performance of a specific device with a realized thickness
mismatch $\beta$ and Mauguin non-ideality $\alpha$. We now use this
expression to derive the figures of merit (FoM) defined in Sec.~\ref{subsec:Figures-of-Merit}.
These FoMs are not a description of one specific device, but rather
a worst-case estimate of performance for a device manufactured with
a maximum thickness inaccuracy $b$ (where $b\ge|\beta|$). For a
specific device, the amplitude modulation is defined in Eq.~(\ref{eq:mod_T_exact-1}).
To estimate the worst-case chromatic loss ($\mathcal{A}^{\left(0\right)}$)
for a device built with tolerance $b$, we evaluate the expression
at the bound $b$. 
\begin{equation}
\mathcal{A}^{\left(0\right)}=2\sin^{2}\left(\frac{1}{2}b\Phi_{s}\right)+2\alpha^{2}\label{eq:A1}
\end{equation}
Here, it is understood that $\Phi_{s}$ represents the cumulative
retardation, and the formula provides the upper bound of distortion.

The phase distortion for a specific device is 
\begin{equation}
\delta\varphi\approx-\varphi\frac{\sin\left(\Phi_{s}\right)}{\Phi_{s}\cos\left(\beta\Phi_{s}\right)}\label{eq:d_fi}
\end{equation}
As this distortion is directly proportional to $\varphi$, the Constant
Phase Distortion is zero in this approximation: $\mathcal{P}_{0}^{\left(0\right)}=0$.
To estimate the worst-case angular phase distortion ($\mathcal{P}_{1}^{\left(0\right)}$),
we again evaluate the function at the tolerance bound $b$ and assume
the worst-case scenario for the oscillating terms within the bandwidth:
\begin{equation}
\mathcal{P}_{1}^{\left(0\right)}=\frac{1}{\Phi_{s}\left|\cos\left(b\Phi_{s}\right)\right|}\label{eq:P_1}
\end{equation}
This estimate assumes that $|\sin\left(\Phi_{s}\right)|$ can reach
$1$ and that the retardation $\Phi_{s}$ and tolerance $b$ combine
to minimize the denominator across the band. This estimate is valid
under the approximation $\left|\cos\left(b\Phi_{s}\right)\right|\gg\alpha$.

\begin{figure}

\includegraphics[width=0.95\columnwidth]{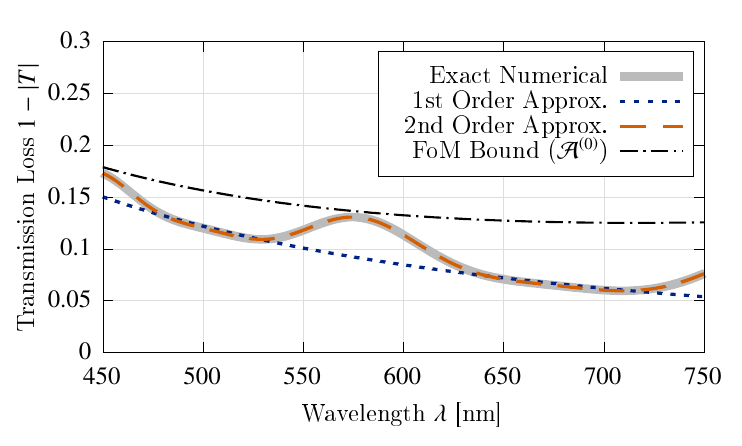}(a)

\includegraphics[width=0.95\columnwidth]{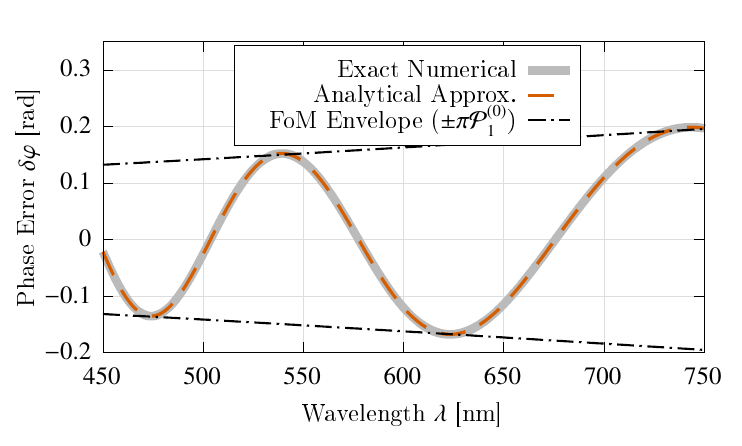}(b)\caption{\label{fig:model_validation}Numerical validation of the theoretical
model for a non-ideal tandem cell}

$\textbf{(a)}$ Amplitude transmission loss ($1-|T|$). $\textbf{(b)}$
Phase error ($\delta\varphi$). Simulation parameters: single-cell
thickness $h=10\,\mu\text{m}$, $b=0.02$, $\Delta n=0.2$, $\varphi=\pi$. 

\end{figure}

To validate the derived analytical expressions and the proposed figures
of merit, we performed exact numerical simulations using the Jones
matrix method. Figure~\ref{fig:model_validation} presents the comparison
between the exact solution and our approximations for a representative
cell with a thickness of $h=10\,\mu\text{m}$ and a thickness mismatch
of $b=0.02$.

As shown in Fig.~\ref{fig:model_validation}(a), the first-order
approximation (dotted line) correctly captures the spectral trend
of the transmission loss but exhibits a magnitude offset. In contrast,
the second-order approximation (dashed line) is virtually indistinguishable
from the exact numerical curve (solid gray line), confirming that
retaining quadratic terms is essential for precise amplitude evaluation.
The Amplitude FoM ($\mathcal{A}^{\left(0\right)}$), plotted as the
black dash-dotted line, correctly predicts the upper bound of these
losses.

Similarly, Fig.~\ref{fig:model_validation}(b) demonstrates that
the phase error oscillates within the envelope defined by the angular
phase distortion parameter $\mathcal{P}_{1}^{\left(0\right)}$ (dash-dotted
lines). This confirms that our derived metrics provide a reliable
worst-case estimate for the device performance.

\section{Compensation Strategies and Discussion}

\subsection{Mitigation of Non-Idealities: The Compensated System Model}

\begin{figure*}
\includegraphics[width=1\textwidth]{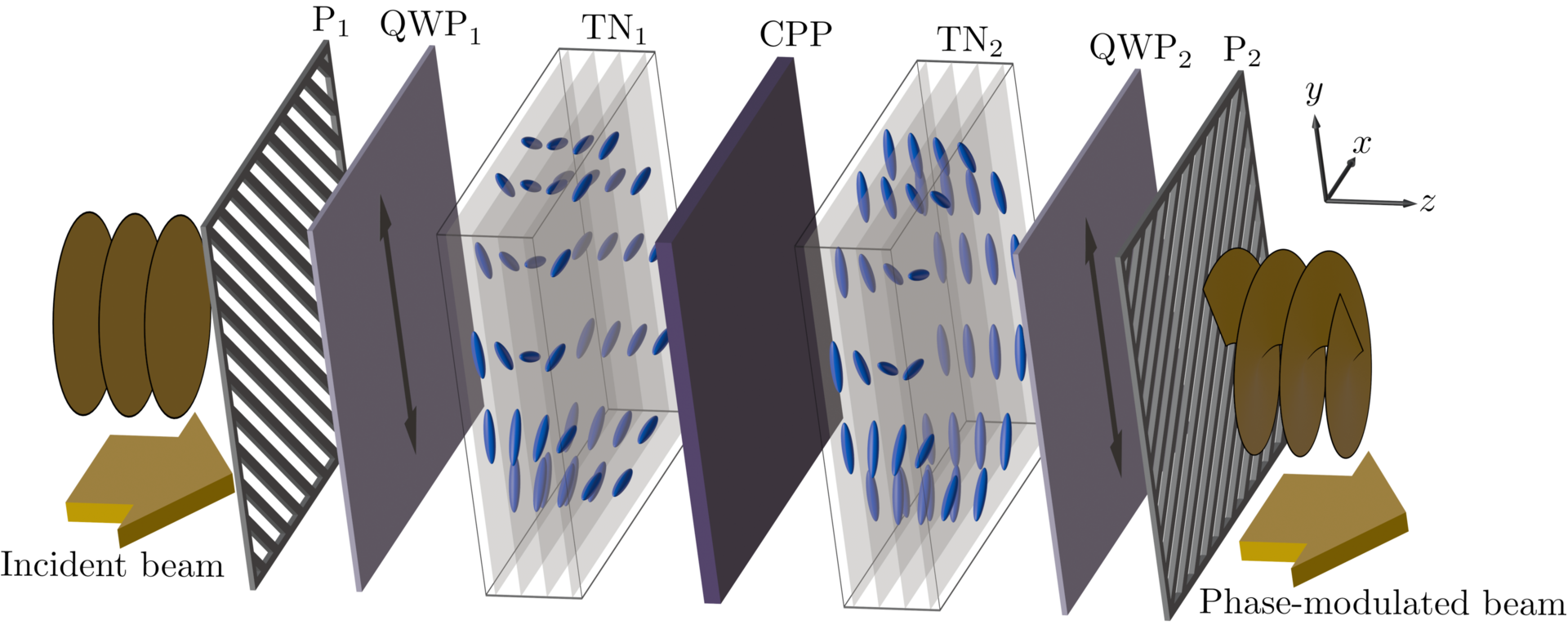}\caption{\label{fig:3}Schematic of the proposed optical setup for the generation
and filtering of optical vortices.}
The incident beam is converted to circular polarization by the input
stage consisting of a linear polarizer ($\text{P}_{1}$, polarization
axis indicated by the parallel grating lines) and an achromatic quarter-wave
plate ($\text{QWP}_{1}$, slow axis indicated by the double-headed
arrow). The tandem cell structure ($\text{TN}_{1}$ and $\text{TN}_{2}$)
imparts the desired phase profile to the beam. A compensating phase
plate ($\text{CPP}$) is inserted at the interface to correct the
system's chromatic response: it is used to nullify manufacturing-induced
amplitude modulation (Strategy A) or to suppress phase distortion
(Strategy B). The output stage, comprising a second quarter-wave plate
($\text{QWP}_{2}$) and an analyzer ($\text{P}_{2}$), filters out
the unwanted parasitic polarization component (the component with
inverted handedness and phase), transmitting only the vortex beam.
\end{figure*}

As established in Sec.~\ref{sec:Analysis-of-a-TNC}, the non-ideal
tandem cell's performance is governed by a fundamental manufacturing
trade-off. Minimizing the angular phase distortion ($\mathcal{P}_{1}$)
requires a large cumulative retardation $\Phi_{s}$. However, a large
$\Phi_{s}$ simultaneously amplifies the system's sensitivity to manufacturing
tolerances, leading to severe amplitude non-achromaticity ($\mathcal{A}$).
This distortion scales with the maximum thickness inaccuracy ($b$)
and cumulative retardation $\Phi_{s}$ as shown in Eq.~(\ref{eq:A1}).
To overcome this limitation and decouple the phase modulation from
the amplitude artifacts, we propose a modified device architecture.
While the director orientations at the external boundaries of the
tandem cell depend on the overall twist angle, the orientations at
the interface between the two cells are fixed (orthogonally along
the $x$ and $y$ axes). A tunable phase retarder (e.g., a voltage-controlled
liquid crystal cell) can therefore be inserted at this interface (depicted
as CPP in Fig.~\ref{fig:3}). This element introduces a controllable
retardation between the $x$ and $y$ polarizations.

The calibration procedure is straightforward: by adjusting the voltage
on the compensation cell to maximize the total light transmission
through the entire optical train (including the final analyzer), one
can actively nullify the effect of $\Phi_{d}$. This active compensation
effectively eliminates the dominant source of amplitude modulation
(the $\cos\Phi_{d}$ term). While minor second-order effects related
to the finite Mauguin parameter persist (as detailed in Appendix \ref{subsec:Supplementary_3}),
the compensated tandem cell achieves a highly achromatic amplitude
response. Its overall transformation is described by:

\begin{equation}
\begin{aligned}T_{\text{CTN}}^{'}= & \exp\left(i\varphi\right)\cos\left(\Phi_{d}+\Phi_{c}\right)-\\
- & i\alpha\exp\left(i\varphi\right)\cos\left(\Phi_{c}\right)\sin\left(\Phi_{s}\right)
\end{aligned}
\label{eq:T_}
\end{equation}

This expression represents the dominant terms derived from the rigorous
analysis. While Eq.~(\ref{eq:T_}) corresponds to the first-order
approximation in terms of the small parameters $\alpha$ and $\beta$,
the complete derivation utilizing the second-order approximation is
provided in Appendix \ref{subsec:Supplementary_3}. The new parameter,
$\Phi_{c}$, which appears in this equation, represents the controllable
phase retardation introduced by the compensating element.

For this derivation, the compensator was assumed to be oriented with
its slow axis along the x-direction (aligned with the director orientation
at the output of the first TN cell) and its fast axis along the $y$-direction.
Thus,$\Phi_{c}$ is specifically defined as the phase lag of the $x$-polarized
component relative to the $y$-polarized component. In the limit where
$\left|\cos\left(\Phi_{d}+\Phi_{c}\right)\right|\gg\alpha$, we can
rewrite  Eq.~(\ref{eq:T_}) in the following form:
\begin{equation}
\begin{aligned}T_{\text{CTN}}^{'}= & \cos\left(\Phi_{d}+\Phi_{c}\right)\times\\
\times & \exp\left(i\varphi\left[1-\frac{\sin\left(\Phi_{s}\right)\cos\left(\Phi_{c}\right)}{\Phi_{s}\cos\left(\Phi_{d}+\Phi_{c}\right)}\right]\right)
\end{aligned}
\label{eq:T_-1}
\end{equation}

As established, the performance of the tandem cell is limited by both
amplitude and phase distortions. With the general theoretical framework
established in Eq.~(\ref{eq:T_-1}), we can now identify specific
physical implementations of the compensation retardation $\Phi_{c}$.
We present these as a hierarchy of strategies, ranging from a rigorous
active control scheme to simplified passive designs. We begin by analyzing
the most comprehensive approach --- active compensation --- which
theoretically allows for the complete elimination of manufacturing
non-idealities and the achievement of perfect broadband performance.

\subsection{Strategy A: Active Compensation for Full Achromaticity }

The first approach targets a complete, broadband compensation of the
amplitude modulation. It involves using a tunable retarder as the
compensating element. Through an active calibration process (e.g.,
maximizing light transmission), the compensator's retardation, $\Phi_{c}$,
is adjusted to satisfy the condition $\Phi_{d}+\Phi_{c}=0$. This
effectively nullifies the manufacturing mismatch $\Phi_{d}$ across
the spectrum.

Evaluating this strategy using the figures of merit defined in Sec.~\ref{subsec:Figures-of-Merit},and
referring to the rigorous second-order analysis in Appendix \ref{subsec:Supplementary_3},
we find that the active compensation theoretically eliminates chromatic
power loss: 

\begin{equation}
\mathcal{A}^{\left(\text{A}\right)}=2\alpha^{2}
\end{equation}

This result indicates that while the compensator corrects for the
retardation errors ($\Phi_{d}$), the fundamental scattering losses
into the orthogonal polarization mode (proportional to $\alpha^{2}$)
persist. However, since $\alpha\propto1/\Phi_{s}$, these losses diminish
rapidly for thicker cells ($\mathcal{A}^{\left(\text{A}\right)}\propto\Phi_{s}^{-2}$),
confirming the high efficiency of the approach.

The phase fidelity is also maintained. As in the uncompensated case,
the constant phase distortion remains zero ($\mathcal{P}_{0}^{\left(\text{A}\right)}=0$).
However, the angular phase distortion is now determined solely by
the cumulative retardation: 
\begin{equation}
\mathcal{P}_{1}^{\left(\text{A}\right)}=\frac{1}{\Phi_{s}}
\end{equation}

Unlike the uncompensated case, where performance is dictated by manufacturing
tolerances (via the parameter $b$), Strategy A effectively decouples
the amplitude fidelity from fabrication errors. The residual loss
arises purely from the non-adiabatic nature of light propagation in
the twisted medium (the finite Mauguin parameter). Since the non-ideality
parameter scales inversely with the cumulative retardation ($\alpha\approx\varphi/\Phi_{s}$),
these residual second-order effects diminish rapidly with increasing
cell thickness.

In summary, Strategy A offers a pathway to near-ideal performance
where the dominant manufacturing-induced artifacts are nullified,
and the ultimate fidelity is limited only by the optical thickness
of the cells.

\begin{figure}
\includegraphics[width=0.95\columnwidth]{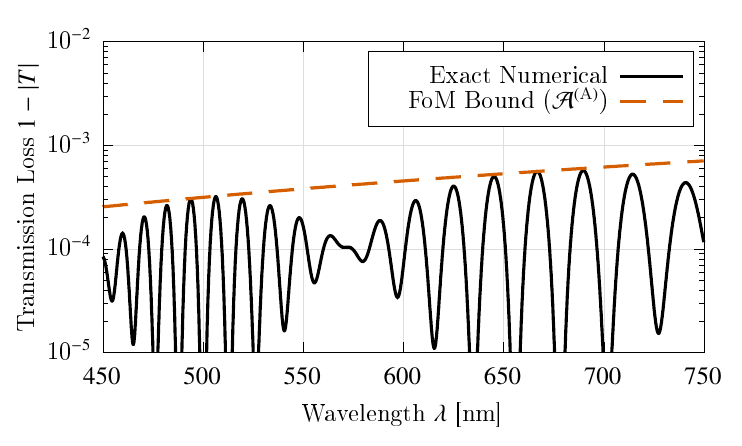}(a)

\includegraphics[width=0.95\columnwidth]{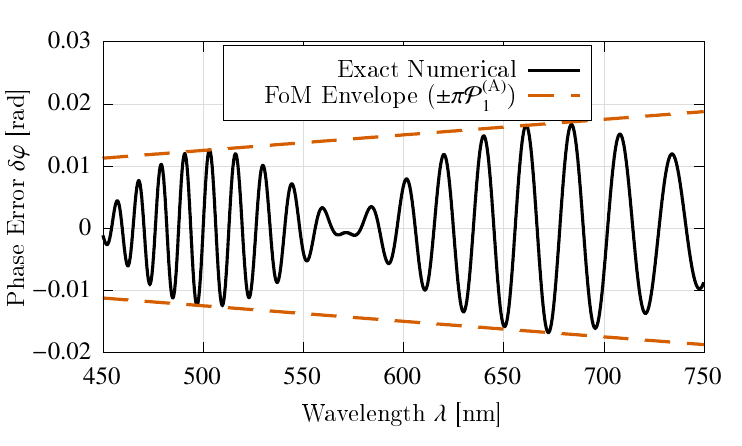}(b)

\caption{\label{fig:Numerical_verification_A}Numerical verification of the
active compensation strategy (Strategy A).}

$\textbf{(a)}$ Residual amplitude transmission loss ($1-|T|$) on
a logarithmic scale. $\textbf{(b)}$ Phase error ($\delta\varphi$).
The simulation confirms that active tuning suppresses manufacturing
artifacts, reducing errors to the fundamental limit defined by second-order
Mauguin scattering ($\mathcal{A}^{\left(\text{A}\right)}=2\alpha^{2}$).
Simulation parameters: single-cell thickness $h=100\,\mu\text{m}$,
$b=0.05$, $\Delta n=0.2$, $\varphi=\pi$.
\end{figure}

To verify the effectiveness of Strategy A, we performed numerical
simulations for an optically thick tandem system ($h=100\,\mu\text{m}$)
subject to a significant manufacturing error ($b=0.05$). The results
are presented in Fig.~\ref{fig:Numerical_verification_A}.

Figure~\ref{fig:Numerical_verification_A}(a) illustrates the residual
amplitude non-achromaticity on a logarithmic scale. The exact numerical
result (solid black line) demonstrates that the active compensation
successfully nullifies the dominant manufacturing-induced modulation
($\cos\left(b\Phi_{s}\right)$ term), reducing losses to the level
of $10^{-4}$. The residual signal is bounded strictly by the theoretical
limit $\mathcal{A}^{\left(\text{A}\right)}=2\alpha^{2}$ (dashed orange
line), confirming that the performance is limited only by the adiabaticity
of the waveguide. It should be noted that these values represent the
polarization conversion efficiency; static insertion losses such as
Fresnel reflections are excluded from this analysis.

Simultaneously, Fig.~\ref{fig:Numerical_verification_A}(b) confirms
the high phase fidelity. By enabling the use of thick cells without
the penalty of amplitude modulation, the angular phase distortion
is suppressed to negligible levels, remaining well within the theoretical
envelope defined by $\pm\pi\mathcal{P}_{1}^{\left(\text{A}\right)}=\pm\pi/\Phi_{s}$.

\subsection{Strategy B: Passive Compensation with a Quarter-Wave Plate}

An alternative strategy prioritizes design simplicity using a static
(passive) compensator. The core of this approach is to use a non-achromatic
quarter-wave plate (QWP) as the compensating element, which sets the
nominal retardation to $\Phi_{c}\approx\pi/2$ at the central angular
frequency $\omega_{0}$. The primary role of the QWP is to suppress
the phase perturbation term (by setting $\cos\left(\Phi_{c}\right)\approx0$).
This choice imposes a new design constraint: the two TN cells must
be intentionally designed to be non-identical to satisfy the dispersion-compensated
amplitude condition $\Phi_{2}\approx\Phi_{1}+\Phi_{c}$ across a broad
spectral range. 

The device is considered quasi-achromatic because its performance
is constrained by two distinct physical limitations:
\begin{enumerate}
\item \textbf{Residual Phase Error:} The phase perturbation is only perfectly
nullified exactly at $\omega_{0}$. At other frequencies ($\omega\ne\omega_{0}$),
$\cos\left(\Phi_{c}\right)$ will be small but non-zero, leading to
a residual, chromatic phase error.
\item \textbf{Amplitude Distortion:} Due to manufacturing tolerances, the
dispersion-compensated condition ($\Phi_{2}=\Phi_{1}+\Phi_{c}$) can
never be perfectly met, resulting in a slight, wavelength-dependent
amplitude modulation.
\end{enumerate}
These limitations are quantified by the following figures of merit:
\begin{equation}
\mathcal{A}^{\left(\text{B}\right)}=2\sin^{2}\left(\frac{1}{2}b\Phi_{s}\right)+\frac{1}{2}\alpha^{2}+2\alpha^{2}\left|\sin\left(\frac{\pi}{2}\frac{\omega-\omega_{0}}{\omega_{0}}\right)\right|\label{eq:A_B}
\end{equation}
\begin{equation}
\mathcal{P}_{1}^{\left(\text{B}\right)}=\frac{\left|\sin\left(\frac{\pi}{2}\frac{\omega-\omega_{0}}{\omega_{0}}\right)\right|}{\Phi_{s}\left|\cos\left(b\Phi_{s}\right)\right|}\label{eq:P_B}
\end{equation}

As the distortion remains proportional to the desired phase $\varphi$,
the constant phase distortion is zero: $\mathcal{P}_{0}^{\left(\text{B}\right)}=0$.

Ultimately, Strategy B is effective within a spectral range where
both the residual amplitude modulation and the residual phase perturbation
are below an acceptable threshold.

\begin{figure}
\includegraphics[width=0.95\columnwidth]{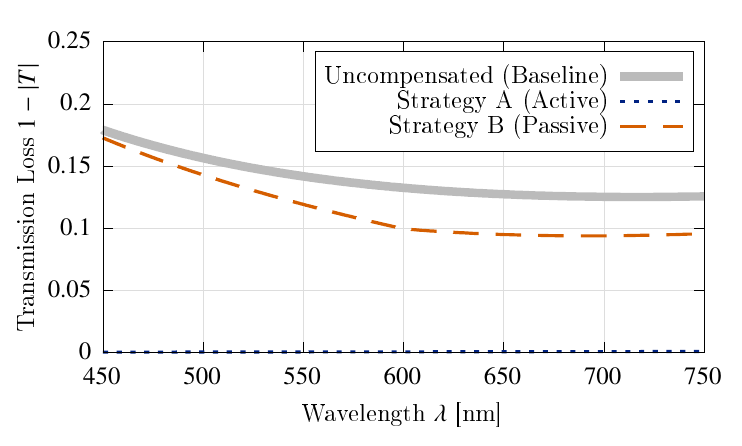}(a)

\includegraphics[width=0.95\columnwidth]{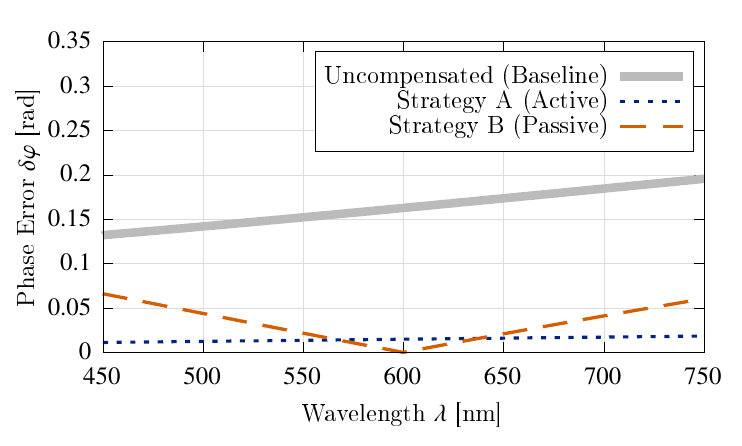}(b)

\caption{\label{fig:Comparative-analysis}Comparative analysis of the figures
of merit for the proposed compensation strategies.}

$\textbf{(a)}$ Amplitude non-achromaticity ($\mathcal{A}$). The
uncompensated baseline $\mathcal{A}^{\left(0\right)}$(solid gray)
and passive strategy $\mathcal{A}^{\left(\text{B}\right)}$ (dashed
line) are of comparable magnitude. The dotted line corresponds to
active strategy $\mathcal{A}^{(\text{A})}$. $\textbf{(b)}$ Angular
phase distortion bounds ($\mathcal{P}=\pi\mathcal{P}_{1}$). The solid
gray line represents the uncompensated baseline ($h=10\,\mu\text{m},b=0.02$).
The dashed line corresponds to passive Strategy B ($h=10\,\mu\text{m},b=0.02$).
The dotted line represents active Strategy A ($h=100\,\mu\text{m}$),
$\Delta n=0.2$.
\end{figure}

The direct quantitative comparison of the strategies is summarized
in Fig.~\ref{fig:Comparative-analysis}.

Figure~\ref{fig:Comparative-analysis}(a) reveals the practical limitation
of passive compensation regarding amplitude fidelity. The curves for
the Uncompensated baseline (gray) and passive Strategy B (dashed line)
overlap significantly. This indicates that for typical manufacturing
tolerances ($b=0.02$ in this simulation), the amplitude non-achromaticity
is dominated by the retardation mismatch term ($\propto\sin^{2}\left(\frac{1}{2}b\Phi_{s}\right)$),
which overshadows the secondary contributions from the Mauguin parameter
or dispersion effects. Consequently, without active tuning to correct
the thickness error, the passive strategy remains vulnerable to fabrication
defects. In sharp contrast, active Strategy A (dotted line) reduces
modulation to negligible levels. Notably, this result is achieved
even with a deliberately relaxed manufacturing tolerance ($b=0.05$)
and a thicker cell, proving that active tuning effectively decouples
amplitude fidelity from fabrication defects.

Figure~\ref{fig:Comparative-analysis}(b) highlights the trade-off
in phase fidelity. Passive Strategy B exhibits a characteristic \textquotedbl V-shape\textquotedbl{}
with a deep minimum at the central wavelength, indicating successful
dispersion matching. However, the error rises at the spectral edges.
Conversely, active Strategy A maintains a consistently low phase distortion
across the entire spectrum. By enabling the use of optically thick
cells ($100\,\mu\text{m}$) without the penalty of amplitude modulation,
this strategy suppresses the fundamental Mauguin parameter $\alpha$,
resulting in superior broadband phase purity.

This comparison underscores a critical practical advantage: the active
compensation mechanism (Strategy A) significantly relaxes manufacturing
tolerances. As illustrated in Fig.~\ref{fig:Comparative-analysis},
the active system suppresses amplitude modulation to negligible levels
even with optically thick cells ($h=100\,\mu\text{m}$) and relaxed
tolerances ($b=0.05$), distinctly outperforming passive designs despite
their tighter constraints ($h=10\,\mu\text{m},b=0.02$).

\subsection{Strategy C: Optimized Operating Point for Narrow-Band Applications}

For applications limited to a relatively narrow spectral bandwidth,
an alternative design approach can be employed. This strategy is applicable
when the cells are manufactured with sufficient precision such that
the retardation mismatch is small ($\Phi_{d}\ll1$). In this regime,
we can exploit the intrinsic properties of the transmission function
to minimize chromatic dispersion without external compensation.

By analyzing the phase term, we observe that the most rapidly oscillating
component is governed by the sinc-like function, $\sin\left(\Phi_{s}\right)/\Phi_{s}$.
The derivative of this function vanishes at its local extrema, implying
that in the vicinity of these points, the phase perturbation is minimally
sensitive to changes in $\Phi_{s}$ (and thus to wavelength). This
allows for a \textquotedbl passive\textquotedbl{} compensation strategy
by engineering the cell thickness such that the cumulative retardation
at the central frequency $\omega_{0}$, denoted as $\Phi_{s0}$, coincides
with a local extremum of the perturbation function.

However, this approach requires a redefinition of the device's modulation
capability. In this regime, the effective phase modulation of the
optical vortex, $\varphi_{\text{mod}}$, is no longer identical to
the geometric twist angle $\varphi$. It is scaled by the perturbation
factor: 
\begin{equation}
\varphi_{\text{mod}}=\varphi\left[1-\frac{\sin\left(\Phi_{s0}\right)}{\Phi_{s0}}\right]\label{eq:fi_mod}
\end{equation}
 To ensure sufficient modulation efficiency (specifically, to ensure
a scaling factor $\ge1$, which enables a full $2\pi$ phase stroke
within the standard director rotation range), we must specifically
target the local minima of the $\text{sinc}$ function.

Evaluating this optimized strategy using our figures of merit (derivation
provided in Appendix \ref{subsec:Derivation-of-Figures}).

The amplitude response is still governed by the thickness manufacturing
tolerance $b$: 
\begin{equation}
\mathcal{A}^{\left(\text{C}\right)}=\frac{1}{2}\left(b\Phi_{s}\right)^{2}+\frac{1}{2}\alpha^{2}\label{eq:A_C}
\end{equation}

Unlike previous strategies, the constant phase distortion is non-zero.
This arises because the optimal retardation condition ($\Phi_{s0}^{\text{opt}}$)
depends on the local twist angle $\varphi$. Since $\varphi$ varies
spatially across the vortex beam while the cell thickness $h$ remains
constant, the optimization condition cannot be perfectly satisfied
at all points simultaneously. However, this error can be minimized
by selecting an optimal design angle $\varphi^{\text{opt}}=\frac{\sqrt{3}}{2}\pi$
(see derivation in Appendix \ref{subsec:Derivation-of-Figures}).

For a design optimized for twist angle ($\varphi^{\text{opt}}=\frac{\sqrt{3}}{2}\pi$),
the constant phase distortion is: 
\begin{equation}
\mathcal{P}_{0}^{\left(\text{C}\right)}=\frac{1}{16\Phi_{s0}}\frac{\Delta\omega}{\omega_{0}}\pi^{3}\label{eq:P0_C}
\end{equation}
 The angular phase distortion $\mathcal{P}_{1}^{\left(\text{C}\right)}$
remains dominated by a second-order quadratic term and a term coupling
the bandwidth with the manufacturing tolerance: 
\begin{equation}
\mathcal{P}_{1}^{\left(\text{C}\right)}=\frac{1}{8}\Phi_{s0}\left(\frac{\Delta\omega}{\omega_{0}}\right)^{2}+\frac{1}{2}b\Phi_{s0}\frac{\Delta\omega}{\omega_{0}}\label{eq:P1_C}
\end{equation}

Here, $\Delta\omega/\omega_{0}$ represents the relative spectral
bandwidth. 

This strategy demonstrates that for narrow-band sources (where $\Delta\omega/\omega_{0}$
is small) and high-precision manufacturing (small $b$), exceptional
phase fidelity can be achieved without any external compensation components,
simply by selecting the correct optical thickness.

\section{Future Outlook and Applications}

\subsection{Static Vortex Generators and Fabrication Challenges}

This architecture holds significant promise for fabricating static,
patterned vortex generators using techniques such as photo-alignment
\citep{Sheremet26012016,C1JM13485J,Schadt_1992}. A single nematic
layer typically allows for a director rotation range restricted to
$-90^{\circ}$ to $+90^{\circ}$ (bounded by disclination lines).
Consequently, the induced phase shift, which in the Mauguin regime
tracks the twist angle, is limited to the same range ($\pi$). This
is insufficient for continuous full-cycle ($2\pi$) phase control
required for integer vortex generation.

Our tandem structure inherently overcomes this limitation. By summing
the rotations of two orthogonal cells, the system extends the available
twist range to $-180^{\circ}$ to $+180^{\circ}$. This provides the
full $2\pi$ phase modulation coverage. The primary challenge in fabricating
such a device lies in the precise registration of the two patterned
substrates. Specifically, the disclination lines --- where the director
twist jumps between the extreme values ($+180^{\circ}$ and $-180^{\circ}$)
--- exist in both cells. To ensure a high-quality phase profile without
artifacts, these disclination lines in the first and second cells
must be perfectly aligned spatially.

\subsection{Towards a Full Complex-Amplitude SLM}

Building upon the static concept, the fundamental principles of the
compensated tandem architecture offer a versatile toolkit for next-generation
dynamic spatial light modulators. By integrating this robust achromatic
phase control with dynamic addressing technologies, such as the OZ-IPS
configuration \citep{OZ_IPShttps://doi.org/10.1002/sdtp.11737,Sato_2020,C6TC05465J,Asagi_2022},
one can envision a universal platform for arbitrary light field synthesis.

The generation of such high-fidelity, structured light fields is a
vibrant area of research with diverse applications ranging from exotic
light-matter interactions to complex optical information processing
\citep{fphy_2021_Plutenko_Vasnetsov,Rubinsztein-Dunlop_2017,Rosales-Guzman_2018}.
The compensated tandem architecture contributes a crucial capability
to this domain: the potential for fully achromatic operation.

However, the practical realization of such a device faces specific
engineering challenges. First, current in-plane switching technologies
(like OZ-IPS) may not provide the full rotation range required for
a complete $2\pi$ phase stroke in a single tandem stack. Consequently,
a \textquotedbl tandem of tandems\textquotedbl{} architecture, comprising
four TN cells, might be necessary to achieve full-cycle modulation.
Second, a fundamental trade-off exists between achromaticity and switching
speed: minimizing phase distortion requires optically thick cells
(large $\Phi_{s}$), which inherently increases the liquid crystal
response time.

\section{Conclusion}

In this work, we have presented a comprehensive analysis of a tandem
twisted nematic (TN) liquid crystal cell as a platform for generating
optical vortices. By developing a rigorous theoretical model that
accounts for key non-idealities --- specifically deviations from
the Mauguin regime and manufacturing imperfections --- we have proposed
and analyzed three distinct compensation strategies that elevate this
device's performance far beyond its conventional applications.

Our primary finding is the development of an active compensation strategy
(Strategy A), which utilizes a tunable retarder to completely nullify
chromatic amplitude modulation. This presents a significant advantage
over existing methods, such as those using q-plates, which inherently
suffer from parasitic intensity modulation despite having an achromatic
phase response. Our approach achieves high efficiency with a virtually
achromatic amplitude response, making it an exceptional candidate
for generating high-quality ``white'' optical vortices.

We also introduced two passive alternatives for specific use cases.
Strategy B offers a simplified design using a static quarter-wave
plate. While quasi-achromatic, its performance is robust, limited
primarily by the precision of manufacturing (thickness tolerance $b$).
Strategy C provides a solution for narrow-band applications without
requiring any additional optical elements. By optimizing the cell's
optical thickness to operate at the local extrema of the phase perturbation
function, this method minimizes chromatic dispersion intrinsically.

In conclusion, the compensated tandem TN cell architecture represents
a powerful and versatile platform. Its ability to achieve near-perfect
phase modulation with controllable amplitude effects makes it a superior
alternative to existing technologies for the generation of optical
vortices and opens exciting avenues for the future development of
next-generation achromatic phase modulators.

\section{Acknowledgments }

This research was supported in part by the grant of National Academy
of Sciences of Ukraine \textquotedblleft Formation and study of optical
beams resistant to atmospheric disturbances for the transmission of
optical information and detection of objects in space\textquotedblright .

\section{Data Availability}

The numerical simulation data underlying the results presented in
this paper are available in \citep{dataset_github}

\bibliographystyle{apsrev4-1}
\bibliography{bib/bibliography}

\appendix

\section{\label{subsec:Supplementary_1}Derivation of the Jones Matrix for
a Single Twisted Nematic Layer Eq.~(\ref{eq:T_TN})}

We start from the differential equation describing the evolution of
the Jones vector in the circular basis, as given by Eq. \ref{eq:DifE}
in the main text:
\begin{equation}
\frac{d}{dz}\boldsymbol{\mathcal{E}}\left(z\right)=i\gamma\begin{pmatrix}0 & \exp\left(i2\phi\right)\\
\exp\left(-i2\phi\right) & 0
\end{pmatrix}\boldsymbol{\mathcal{E}}\left(z\right)\label{eq:DifE-1}
\end{equation}
For the case of a linear director twist, where, $\phi\left(z\right)=\phi_{0}+qz$,
we can rewrite this matrix equation as a system of two coupled, first-order
differential equations for the components $\mathcal{E}_{1}$ and $\mathcal{E}_{2}$
\begin{equation}
\begin{cases}
\frac{d}{dz}\mathcal{E}_{1}=i\gamma\exp\left(i2\left(\phi_{0}+qz\right)\right)\mathcal{E}_{2}\\
\frac{d}{dz}\mathcal{E}_{2}=i\gamma\exp\left(-i2\left(\phi_{0}+qz\right)\right)\mathcal{E}_{1}
\end{cases}\label{eq:system}
\end{equation}
The standard procedure to solve this system is to decouple these equations
to obtain a single second-order linear homogeneous differential equation
with constant coefficients. By differentiating the first system Eq.~(\ref{eq:system})
and subsequently substituting for all terms containing the variable
$\mathcal{E}_{2}$ using the original system, we obtain the following
second-order differential equation with constant coefficients:

\begin{equation}
\frac{d^{2}}{dz^{2}}\mathcal{E}_{1}-2qi\frac{d}{dz}\mathcal{E}_{1}+\gamma^{2}\mathcal{E}_{1}=0\label{eq:d_dz}
\end{equation}
\\
The solution of this equation in general case is 
\begin{equation}
\begin{aligned}\mathcal{E}_{1}\left(z\right)= & A\exp\left(i\left(q+\gamma^{'}\right)z\right)+\\
+ & B\exp\left(i\left(q-\gamma^{'}\right)z\right)
\end{aligned}
\label{eq:E1}
\end{equation}
where $A$ and $B$ are complex constants determined by the initial
conditions, and $\gamma^{'}$ is defined as
\begin{equation}
\gamma^{'}=\sqrt{\gamma^{2}+q^{2}}
\end{equation}
The second component, $\mathcal{E}_{2}\left(z\right)$, can then be
found by substituting the solution for $\mathcal{E}_{1}\left(z\right)$
and its derivative back into the first equation of the original system
Eq.~(\ref{eq:system}). After performing the substitution and simplification,
we obtain

\begin{equation}
\begin{aligned}\mathcal{E}_{2}\left(z\right)= & \frac{q+\gamma^{'}}{\gamma}A\exp\left(-i\left[qz-\gamma^{'}z+2\phi_{0}\right]\right)+\\
+ & \frac{q-\gamma^{'}}{\gamma}B\exp\left(-i\left[qz+\gamma^{'}z+2\phi_{0}\right]\right)
\end{aligned}
\label{eq:E2}
\end{equation}
To find the specific solution for a given input polarization, we need
to express the complex constants $A$ and $B$ in terms of the initial
conditions of the field at $z=0$, which we denote as $\boldsymbol{\mathcal{E}}\left(0\right)=\left(\mathcal{E}_{1}\left(0\right),\mathcal{E}_{2}\left(0\right)\right)^{T}$.
By setting $z=0$ in Eqs.~(\ref{eq:E1})--(\ref{eq:E2}), we obtain
the following system of linear equations for $A$ and $B$:
\begin{equation}
\begin{gathered}\begin{cases}
\mathcal{E}_{1}\left(0\right)= & A+B\\
\mathcal{E}_{2}\left(0\right)= & \frac{q+\gamma^{'}}{\gamma}A\exp\left(-i2\phi_{0}\right)+\\
 & +\frac{q-\gamma^{'}}{\gamma}B\exp\left(-i2\phi_{0}\right)
\end{cases}\end{gathered}
\label{eq:E_0__E_1}
\end{equation}
Solving this system yields the expressions for $A$ and $B$

\begin{equation}
A=\frac{\gamma^{'}-q}{2\gamma^{'}}\mathcal{E}_{1}\left(0\right)+\frac{\gamma}{2\gamma^{'}}\exp\left(i2\phi_{0}\right)\mathcal{E}_{2}\left(0\right)\label{eq:A__}
\end{equation}
\begin{equation}
B=\frac{q+\gamma^{'}}{2\gamma^{'}}\mathcal{E}_{1}\left(0\right)-\frac{\gamma}{2\gamma^{'}}\exp\left(i2\phi_{0}\right)\mathcal{E}_{2}\left(0\right)\label{eq:B__}
\end{equation}
By substituting these expressions for $A$ and $B$ back into the
general solutions Eqs.~(\ref{eq:E1})--(\ref{eq:E2}), we can construct
the Jones matrix that maps the input state $\boldsymbol{\mathcal{E}}\left(0\right)$
to the output state $\boldsymbol{\mathcal{E}}\left(h\right)$

\begin{widetext}

\begin{equation}
\boldsymbol{\mathcal{E}}\left(h\right)=\begin{pmatrix}\exp\left(iqh\right)\left[\cos\left(\gamma^{'}h\right)-i\frac{q}{\gamma^{'}}\sin\left(\gamma^{'}h\right)\right] & i\frac{\gamma}{\gamma^{'}}\exp\left(i\left(2\phi_{0}+qh\right)\right)\sin\left(\gamma^{'}h\right)\\
i\frac{\gamma}{\gamma^{'}}\exp\left(-i\left(2\phi_{0}+qh\right)\right)\sin\left(\gamma^{'}h\right) & \exp\left(-iqh\right)\left[\cos\left(\gamma^{'}h\right)+i\frac{q}{\gamma^{'}}\sin\left(\gamma^{'}h\right)\right]
\end{pmatrix}\boldsymbol{\mathcal{E}}\left(0\right)\label{eq:E_h}
\end{equation}
\end{widetext}

\section{\label{subsec:Supplementary_2}Approximation of the Jones Matrix
for the Non-Ideal Tandem Cell Eq.~(\ref{eq:TE_CTN2})}

The total Jones matrix for the non-ideal tandem system, $T$ , is
found by multiplying the individual Jones matrices of the first and
second cells, $T_{\text{TN1}}^{\mathcal{E}}\left(h_{1}\right)$ and
$T_{\text{TN2}}^{\mathcal{E}}\left(h_{2}\right)$ respectively:
\begin{equation}
T=T_{\text{TN2}}^{\mathcal{E}}\left(h_{2}\right)T_{\text{TN1}}^{\mathcal{E}}\left(h_{1}\right)
\end{equation}

The parameters for each cell correspond to those in Eq.~(\ref{eq:T_TN})
of the main text, but are now denoted with subscripts $1$ and $2$
to distinguish between the two cells. We assume the same liquid crystal
material is used in both cells, so the anisotropy parameter $\gamma$
is identical for both. The director twist in each cell is also set
to be the same, with a total twist of $\varphi$ for the system, meaning
$\phi_{d1}=\phi_{d2}=\varphi/2$.

The specific orientation of the tandem system --- where the director
at the output of the first cell is along the $x$-axis and the director
at the input of the second cell is along the $y$-axis --- is achieved
by setting the mean director angles as follows: for the first cell,
we set $\bar{\phi}_{1}=-\varphi/4$, and for the second cell $\bar{\phi}_{2}=\pi/2+\varphi/4$ 

\begin{widetext}
\begin{equation}
T_{\text{TN1}}^{\mathcal{E}}\left(h_{1}\right)=\begin{pmatrix}\exp\left(i\frac{\varphi}{2}\right)\left[\cos\left(\gamma_{1}^{'}h_{1}\right)-i\frac{q_{1}}{\gamma_{1}^{'}}\sin\left(\gamma_{1}^{'}h_{1}\right)\right] & i\frac{\gamma}{\gamma_{1}^{'}}\exp\left(-i\frac{\varphi}{2}\right)\sin\left(\gamma_{1}^{'}h_{1}\right)\\
i\frac{\gamma}{\gamma_{1}^{'}}\exp\left(i\frac{\varphi}{2}\right)\sin\left(\gamma_{1}^{'}h_{1}\right) & \exp\left(-i\frac{\varphi}{2}\right)\left[\cos\left(\gamma_{1}^{'}h_{1}\right)+i\frac{q_{1}}{\gamma_{1}^{'}}\sin\left(\gamma_{1}^{'}h_{1}\right)\right]
\end{pmatrix}\label{eq:T_E_TN1}
\end{equation}
\begin{equation}
T_{\text{TN2}}^{\mathcal{E}}\left(h_{2}\right)=\begin{pmatrix}\exp\left(i\frac{\varphi}{2}\right)\left[\cos\left(\gamma_{2}^{'}h_{2}\right)-i\frac{q_{2}}{\gamma_{2}^{'}}\sin\left(\gamma_{2}^{'}h_{2}\right)\right] & -i\frac{\gamma}{\gamma_{2}^{'}}\exp\left(i\frac{\varphi}{2}\right)\sin\left(\gamma_{2}^{'}h_{2}\right)\\
-i\frac{\gamma}{\gamma_{2}^{'}}\exp\left(-i\frac{\varphi}{2}\right)\sin\left(\gamma_{2}^{'}h_{2}\right) & \exp\left(-i\frac{\varphi}{2}\right)\left[\cos\left(\gamma_{2}^{'}h_{2}\right)+i\frac{q_{2}}{\gamma_{2}^{'}}\sin\left(\gamma_{2}^{'}h_{2}\right)\right]
\end{pmatrix}\label{eq:T_E_TN2}
\end{equation}

\end{widetext}

The resulting matrix $T$ is algebraically complex. For clarity, instead
of presenting the full $2\times2$ matrix at once, we will list each
of its components individually. 

\begin{widetext}
\begin{equation}
T_{11}=\exp\left(i\varphi\right)\left\{ \left[\cos\left(\gamma_{2}^{'}h_{2}\right)-i\frac{q_{2}}{\gamma_{2}^{'}}\sin\left(\gamma_{2}^{'}h_{2}\right)\right]\left[\cos\left(\gamma_{1}^{'}h_{1}\right)-i\frac{q_{1}}{\gamma_{1}^{'}}\sin\left(\gamma_{1}^{'}h_{1}\right)\right]+\frac{\gamma^{2}}{\gamma_{2}^{'}\gamma_{1}^{'}}\sin\left(\gamma_{2}^{'}h_{2}\right)\sin\left(\gamma_{1}^{'}h_{1}\right)\right\} \label{eq:T11}
\end{equation}
\begin{equation}
T_{12}=i\frac{\gamma}{\gamma_{1}^{'}}\left[\cos\left(\gamma_{2}^{'}h_{2}\right)-i\frac{q_{2}}{\gamma_{2}^{'}}\sin\left(\gamma_{2}^{'}h_{2}\right)\right]\sin\left(\gamma_{1}^{'}h_{1}\right)-i\frac{\gamma}{\gamma_{2}^{'}}\sin\left(\gamma_{2}^{'}h_{2}\right)\left[\cos\left(\gamma_{1}^{'}h_{1}\right)+i\frac{q_{1}}{\gamma_{1}^{'}}\sin\left(\gamma_{1}^{'}h_{1}\right)\right]\label{eq:T12}
\end{equation}
\begin{equation}
T_{21}=-i\frac{\gamma}{\gamma_{2}^{'}}\sin\left(\gamma_{2}^{'}h_{2}\right)\left[\cos\left(\gamma_{1}^{'}h_{1}\right)-i\frac{q_{1}}{\gamma_{1}^{'}}\sin\left(\gamma_{1}^{'}h_{1}\right)\right]+i\frac{\gamma}{\gamma_{1}^{'}}\left[\cos\left(\gamma_{2}^{'}h_{2}\right)+i\frac{q_{2}}{\gamma_{2}^{'}}\sin\left(\gamma_{2}^{'}h_{2}\right)\right]\sin\left(\gamma_{1}^{'}h_{1}\right)\label{eq:T21}
\end{equation}
\begin{equation}
T_{22}=\exp\left(-i\varphi\right)\left\{ \frac{\gamma^{2}}{\gamma_{1}^{'}\gamma_{2}^{'}}\sin\left(\gamma_{2}^{'}h_{2}\right)\sin\left(\gamma_{1}^{'}h_{1}\right)+\left[\cos\left(\gamma_{2}^{'}h_{2}\right)+i\frac{q_{2}}{\gamma_{2}^{'}}\sin\left(\gamma_{2}^{'}h_{2}\right)\right]\left[\cos\left(\gamma_{1}^{'}h_{1}\right)+i\frac{q_{1}}{\gamma_{1}^{'}}\sin\left(\gamma_{1}^{'}h_{1}\right)\right]\right\} \label{eq:T22}
\end{equation}

\end{widetext}

The exact expressions for the matrix components Eqs.~(\ref{eq:T11})--(\ref{eq:T22})
are algebraically complex. To facilitate a rigorous analysis, we introduce
a set of independent dimensionless parameters characterizing the system's
non-idealities.

First, we define the thickness mismatch parameter $\beta$ and the
mean cell thickness $h$: 
\begin{equation}
\beta=\frac{h_{1}-h_{2}}{h_{1}+h_{2}},\quad h=\frac{h_{1}+h_{2}}{2}
\end{equation}
 Consequently, the individual thicknesses are expressed as $h_{1}=h\left(1+\beta\right)$
and $h_{2}=h\left(1-\beta\right)$.

Next, we consider the Mauguin parameters for each cell, defined using
the effective wavenumber inside the twisted medium, $\gamma'=\sqrt{\gamma^{2}+q^{2}}$.
Thus, $\alpha_{1}=q_{1}/\gamma'_{1}$ and $\alpha_{2}=q_{2}/\gamma'_{2}$.
We introduce the mean Mauguin parameter $\alpha$: 
\begin{equation}
\alpha=\frac{1}{2}\left(\alpha_{1}+\alpha_{2}\right)
\end{equation}
 To derive the relationship between the individual parameters $\alpha_{1,2}$
and the independent variables $\alpha$ and $\beta$, we analyze the
difference $\alpha_{2}-\alpha_{1}$. Consistent with our overall perturbative
approach, we explicitly limit our analysis to the second order of
smallness with respect to the parameters $\alpha$ and $\beta$. Thus,
\begin{equation}
\alpha_{2}-\alpha_{1}=\frac{q_{2}}{\gamma_{2}^{'}}-\frac{q_{1}}{\gamma_{1}^{'}}=2\alpha\frac{q_{2}\gamma_{1}^{'}-q_{1}\gamma_{2}^{'}}{q_{1}\gamma_{2}^{'}+q_{2}\gamma_{1}^{'}}
\end{equation}

Under this constraint (neglecting cubic terms and higher), we have:
\begin{equation}
\alpha_{2}-\alpha_{1}=2\alpha\frac{h_{2}^{-1}-h_{1}^{-1}}{h_{2}^{-1}+h_{1}^{-1}}=2\alpha\beta
\end{equation}

From this relation, we derive the expressions for the individual parameters:
\begin{equation}
\alpha_{1}=\alpha\left(1-\beta\right),\quad\alpha_{2}=\alpha\left(1+\beta\right)
\end{equation}
 We now approximate the Jones matrix components by expanding their
coefficients in powers of $\alpha$ and $\beta$, retaining terms
up to the second order (including $\alpha^{2}$, $\beta^{2}$ and
$\alpha\beta$). The arguments of the trigonometric functions (like
$\gamma^{'}h$) are left unexpanded.

Following this methodology, the approximated matrix components are
given by:

\begin{widetext}

\begin{equation}
\begin{aligned}T_{11}= & \exp\left(i\varphi\right)\left\{ \cos\left(\gamma_{1}^{'}h_{1}-\gamma_{2}^{'}h_{2}\right)-i\alpha\sin\left(\gamma_{2}^{'}h_{2}+\gamma_{1}^{'}h_{1}\right)\right\} \\
- & \exp\left(i\varphi\right)\left\{ 2\alpha^{2}\sin\left(\gamma_{1}^{'}h_{1}\right)\sin\left(\gamma_{2}^{'}h_{2}\right)+i\alpha\beta\sin\left(\gamma_{2}^{'}h_{2}-\gamma_{1}^{'}h_{1}\right)\right\} 
\end{aligned}
\label{eq:T11-2}
\end{equation}
\begin{equation}
T_{12}=i\left(1-\frac{1}{2}\alpha^{2}\right)\sin\left(\gamma_{1}^{'}h_{1}-\gamma_{2}^{'}h_{2}\right)+2\alpha\sin\left(\gamma_{1}^{'}h_{1}\right)\sin\left(\gamma_{2}^{'}h_{2}\right)\label{eq:T12-2}
\end{equation}
\begin{equation}
T_{21}=i\left(1-\frac{1}{2}\alpha^{2}\right)\sin\left(\gamma_{1}^{'}h_{1}-\gamma_{2}^{'}h_{2}\right)-2\alpha\sin\left(\gamma_{2}^{'}h_{2}\right)\sin\left(\gamma_{1}^{'}h_{1}\right)\label{eq:T21-2}
\end{equation}
\begin{equation}
\begin{aligned}T_{22}= & \exp\left(-i\varphi\right)\left\{ \cos\left(\gamma_{1}^{'}h_{1}-\gamma_{2}^{'}h_{2}\right)+i\alpha\sin\left(\gamma_{2}^{'}h_{2}+\gamma_{1}^{'}h_{1}\right)\right\} -\\
- & \exp\left(-i\varphi\right)\left\{ 2\alpha^{2}\sin\left(\gamma_{2}^{'}h_{2}\right)\sin\left(\gamma_{1}^{'}h_{1}\right)-i\alpha\beta\sin\left(\gamma_{2}^{'}h_{2}-\gamma_{1}^{'}h_{1}\right)\right\} 
\end{aligned}
\label{eq:T22-2}
\end{equation}

\end{widetext}

Finally, for the filtered optical system where only the circular polarization
component is preserved, the total transmission coefficient $T_{\text{system}}$
is determined primarily by $T_{11}$. The derived expression in Eqs.~(\ref{eq:T11-2})--(\ref{eq:T22-2})
provides the basis for the figures of merit analysis, capturing both
the first-order phase distortions and the second-order amplitude modulation
effects.

\section{\label{subsec:Supplementary_3}Derivation of the Transmission Coefficient
for the Filtered System (Eq. \ref{eq:T_CTN})}

To derive the transmission coefficient for the complete filtered system,
we first write the Jones matrix for the central tandem element, which
now includes the compensating plate between the two TN cells. This
matrix, $T_{\text{core}}$ , is the product of the three individual
matrices:
\begin{equation}
T_{\text{core}}=T_{\text{TN2}}^{\mathcal{E}}\left(h_{2}\right)T_{\text{C}}^{\mathcal{E}}\left(\Phi_{c}\right)T_{\text{TN1}}^{\mathcal{E}}\left(h_{1}\right)\label{eq:T_core}
\end{equation}

The matrices $T_{\text{TN1}}^{\mathcal{E}}\left(h_{1}\right)$ and
$T_{\text{TN2}}^{\mathcal{E}}\left(h_{2}\right)$ are the Jones matrices
for the two twisted nematic cells, with components as derived in the
previous section. The central matrix, $T_{\text{C}}^{\mathcal{E}}\left(\Phi_{c}\right)$,
represents the compensating phase plate. For a linear retarder with
its slow axis oriented along the $x$-direction and a controllable
retardation of $\Phi_{c}$, this matrix in the circular basis is:
\begin{equation}
T_{\text{C}}^{\mathcal{E}}\left(\Phi_{c}\right)=\begin{pmatrix}\cos\left(\Phi_{c}\right) & i\sin\left(\Phi_{c}\right)\\
i\sin\left(\Phi_{c}\right) & \cos\left(\Phi_{c}\right)
\end{pmatrix}\label{eq:TE2-1}
\end{equation}
We can now derive the total transmission coefficient, $T_{\text{system}}$,
for the full optical train. The first two elements, a linear polarizer
and a quarter-wave plate, prepare a pure right-hand circular (RHC)
input state, which is described by the Jones vector $\left(1,0\right)^{T}$
in the circular basis.

Correspondingly, the final two elements (the second QWP and the analyzer)
are configured to act as a filter that transmits only the RHC component
of the light.

Therefore, the complex transmission coefficient for the entire system
is simply the $\left(1,1\right)$ element of the core Jones matrix,
$T_{\text{core}}$ , defined in Eq.~(\ref{eq:T_core}). This simplifies
the calculation significantly, as we do not need to compute the other
three matrix elements. The expression for this coefficient, $T_{\text{system}}=\left(T_{\text{core}}\right)_{11}$,
is obtained by performing the matrix multiplication $\left(T_{\text{TN2}}^{\mathcal{E}}T_{\text{C}}^{\mathcal{E}}T_{\text{TN1}}^{\mathcal{E}}\right)_{11}$,
which yields:

\begin{equation}
\begin{aligned}T_{\text{system}}= & \left(T_{\text{TN2}}^{\mathcal{E}}\right)_{11}\left(T_{\text{C}}^{\mathcal{E}}\right)_{11}\left(T_{\text{TN1}}^{\mathcal{E}}\right)_{11}+\\
+ & \left(T_{\text{TN2}}^{\mathcal{E}}\right)_{11}\left(T_{\text{C}}^{\mathcal{E}}\right)_{12}\left(T_{\text{TN1}}^{\mathcal{E}}\right)_{21}+\\
+ & \left(T_{\text{TN2}}^{\mathcal{E}}\right)_{12}\left(T_{\text{C}}^{\mathcal{E}}\right)_{21}\left(T_{\text{TN1}}^{\mathcal{E}}\right)_{11}\\
+ & \left(T_{\text{TN2}}^{\mathcal{E}}\right)_{12}\left(T_{\text{C}}^{\mathcal{E}}\right)_{22}\left(T_{\text{TN1}}^{\mathcal{E}}\right)_{21}
\end{aligned}
\end{equation}

After substituting the corresponding elements of the matrices $T_{\text{TN1}}^{\mathcal{E}}$,
$T_{\text{C}}^{\mathcal{E}}$ and $T_{\text{TN2}}^{\mathcal{E}}$
into the expression for $T_{\text{system}}=\left(T_{\text{core}}\right)_{11}$,
we obtain the following exact expression for the transmission coefficient:

\begin{widetext}
\begin{equation}
\begin{aligned}T_{\text{system}}= & \exp\left(i\varphi\right)\left[\cos\left(\gamma_{2}^{'}h_{2}\right)-i\frac{q_{2}}{\gamma_{2}^{'}}\sin\left(\gamma_{2}^{'}h_{2}\right)\right]\times\\
\times & \left\{ \cos\left(\Phi_{c}\right)\left[\cos\left(\gamma_{1}^{'}h_{1}\right)-i\frac{q_{1}}{\gamma_{1}^{'}}\sin\left(\gamma_{1}^{'}h_{1}\right)\right]-\frac{\gamma}{\gamma_{1}^{'}}\sin\left(\Phi_{c}\right)\sin\left(\gamma_{1}^{'}h_{1}\right)\right\} +\\
+ & \frac{\gamma}{\gamma_{2}^{'}}\exp\left(i\varphi\right)\sin\left(\gamma_{2}^{'}h_{2}\right)\left\{ \sin\left(\Phi_{c}\right)\left[\cos\left(\gamma_{1}^{'}h_{1}\right)-i\frac{q_{1}}{\gamma_{1}^{'}}\sin\left(\gamma_{1}^{'}h_{1}\right)\right]+\frac{\gamma}{\gamma_{1}^{'}}\cos\left(\Phi_{c}\right)\sin\left(\gamma_{1}^{'}h_{1}\right)\right\} 
\end{aligned}
\label{eq:T_system_raw}
\end{equation}

\end{widetext}

This exact expression is cumbersome. We now simplify it by applying
the second-order approximation methodology discussed in previous section.
This involves expanding the expression and retaining only the terms
that are not higher than second order in the small parameters $\alpha$
and $\beta$, while keeping the arguments of the trigonometric functions
in their exact form.

After significant algebraic simplification, the transmission coefficient
reduces to the much more compact and insightful form:

\begin{widetext}
\begin{align}
T_{\text{system}} & =\exp\left(i\varphi\right)\left[\cos\left(\Phi_{c}+\gamma_{1}^{'}h_{1}-\gamma_{2}^{'}h_{2}\right)-i\alpha\cos\left(\Phi_{c}\right)\sin\left(\gamma_{2}^{'}h_{2}+\gamma_{1}^{'}h_{1}\right)\right]+\label{eq:T_system}\\
 & +i\alpha\beta\exp\left(i\varphi\right)\left[\sin\left(\gamma_{1}^{'}h_{1}-\gamma_{2}^{'}h_{2}\right)\cos\left(\Phi_{c}\right)+2\sin\left(\gamma_{1}^{'}h_{1}\right)\sin\left(\gamma_{2}^{'}h_{2}\right)\sin\left(\Phi_{c}\right)\right]\nonumber \\
 & +\alpha^{2}\exp\left(i\varphi\right)\left[\frac{1}{2}\sin\left(\gamma_{1}^{'}h_{1}-\gamma_{2}^{'}h_{2}\right)\sin\left(\Phi_{c}\right)-2\sin\left(\gamma_{1}^{'}h_{1}\right)\sin\left(\gamma_{2}^{'}h_{2}\right)\cos\left(\Phi_{c}\right)\right]\nonumber 
\end{align}

\end{widetext}

\section{\label{subsec:Derivation-of-Figures_B}Derivation of Figures of Merit
for Strategy B}

To strictly derive the amplitude fidelity, we refer to the exact transmission
coefficient derived from the Jones matrix multiplication, denoted
here as Eq.~(\ref{eq:T_system_raw}) (referring to the unexpanded
matrix product in Supplementary Materials \ref{subsec:Supplementary_3}).
Unlike the previous derivation for the active strategy, where $\Phi_{c}$
was an independent variable, here $\Phi_{c}$ is determined by the
physical properties of the quarter-wave plate.

We assume the QWP is made of a material with similar dispersion characteristics
to the liquid crystal, or that its dispersion tracks the LC linearly.

\begin{equation}
\Phi_{c}\left(\omega\right)=\frac{\pi}{2}\frac{\omega}{\omega_{0}}
\end{equation}
 where $\omega_{0}$ is the central design frequency at which the
plate acts as a perfect $\lambda/4$ retarder ($\Phi_{c}\left(\omega_{0}\right)=\Phi_{c0}=\pi/2$).

We define the \textquotedbl equivalent thickness\textquotedbl{} of
the compensator, $h_{c}$, as the thickness of a hypothetical LC layer
that would produce the same retardation $\Phi_{c0}=\Phi_{c}\left(\omega_{0}\right)=\pi/2$
at the central frequency $\omega_{0}$: 
\begin{equation}
h_{c}=\Phi_{c0}\gamma^{-1}\left(\omega_{0}\right)=\frac{\pi}{2}\gamma^{-1}\left(\omega_{0}\right)
\end{equation}
 To analyze the system as a unified structure, we redefine the thickness
mismatch parameter $\beta$ to include this effective thickness. The
effective mismatch of the compensated system is: 
\begin{equation}
\beta=\frac{h_{1}-h_{2}+h_{c}}{h_{2}+h_{1}}\label{eq:beta_pi_2}
\end{equation}
 By design, Strategy B aims to satisfy the condition $h_{1}-h_{2}+h_{c}=0$,
which corresponds to $\beta=0$. However, due to manufacturing inaccuracies,
the realized parameter deviates from zero. The maximum thickness inaccuracy
$b$ sets the bound for this deviation: $|\beta|\le b$.

With the redefinition of the thickness mismatch parameter $\beta$
Eq.~(\ref{eq:beta_pi_2}), we must update the expressions for the
geometric and optical parameters used in the perturbative expansion.

We retain the definitions for the mean Mauguin parameter $\alpha$
and the mean cell thickness $h$ as established in Appendix~\ref{subsec:Supplementary_2}:
\begin{equation}
\alpha_{1}=q_{1}/\gamma'_{1},\quad\alpha_{2}=q_{2}/\gamma'_{2},
\end{equation}
\begin{equation}
\alpha=\frac{1}{2}\left(\alpha_{1}+\alpha_{2}\right),\quad h=\frac{h_{1}+h_{2}}{2}
\end{equation}
 However, the expressions for the individual cell thicknesses $h_{1,2}$
must be reformulated to account for the inclusion of the compensator
equivalent thickness $h_{c}$ in the definition of $\beta$. Solving
for $h_{1}$ and $h_{2}$, we obtain: 
\begin{equation}
h_{1}=h\left(1+\beta\right)-\frac{1}{2}h_{c},\quad h_{2}=h\left(1-\beta\right)+\frac{1}{2}h_{c}
\end{equation}
 Consequently, the difference between the Mauguin parameters, which
scales with the physical thickness difference $(h_{1}-h_{2})$, is
also modified. Using the relation $\alpha_{2}-\alpha_{1}\approx2\alpha\left(h_{1}-h_{2}\right)/\left(h_{1}+h_{2}\right)$,
we find: 
\begin{equation}
\alpha_{2}-\alpha_{1}=2\alpha\left(\beta-\frac{1}{2}\frac{h_{c}}{h}\right)
\end{equation}
 This leads to the updated expressions for the individual parameters
$\alpha_{1}$ and $\alpha_{2}$ in terms of the independent variables:
\begin{equation}
\alpha_{1}=\alpha\left(1-\beta+\frac{1}{2}\frac{h_{c}}{h}\right)
\end{equation}
\begin{equation}
\alpha_{2}=\alpha\left(1+\beta-\frac{1}{2}\frac{h_{c}}{h}\right)
\end{equation}
 We retain the standard form of the definitions for the phase arguments,
noting that their values are now governed by the modified thicknesses
derived above: 
\begin{equation}
\Phi_{d}=\gamma_{1}^{'}h_{1}-\gamma_{2}^{'}h_{2},\quad\Phi_{s}=\gamma_{1}^{'}h_{1}+\gamma_{2}^{'}h_{2}
\end{equation}

Substituting these redefined parameters into the exact transmission
coefficient Eq.~(\ref{eq:T_system_raw}) (derived from the raw Jones
matrix multiplication) and performing a series expansion, we retain
terms up to the second order in the small parameters $\alpha$ and
$\beta$.

Using the linear dispersion relation for the QWP, $\Phi_{c}\left(\omega\right)=\frac{\pi}{2}\frac{\omega}{\omega_{0}}$,
the transmission coefficient for Strategy B takes the following form:

\begin{widetext}
\begin{align}
T_{\text{system}} & =\exp\left(i\varphi\right)\left[\cos\left(\frac{\pi}{2}\frac{\omega}{\omega_{0}}+\Phi_{d}\right)-i\alpha\cos\left(\frac{\pi}{2}\frac{\omega}{\omega_{0}}\right)\sin\left(\Phi_{s}\right)\right]+\label{eq:T_system-1-1}\\
 & +i\alpha\left[\beta-\frac{1}{2}\frac{h_{c}}{h}\right]\exp\left(i\varphi\right)\left[\sin\left(\Phi_{d}\right)\cos\left(\frac{\pi}{2}\frac{\omega}{\omega_{0}}\right)+2\sin\left(\gamma_{1}^{'}h_{1}\right)\sin\left(\gamma_{2}^{'}h_{2}\right)\sin\left(\frac{\pi}{2}\frac{\omega}{\omega_{0}}\right)\right]\nonumber \\
 & +\alpha^{2}\exp\left(i\varphi\right)\left[\frac{1}{2}\sin\left(\Phi_{d}\right)\sin\left(\frac{\pi}{2}\frac{\omega}{\omega_{0}}\right)-2\sin\left(\gamma_{1}^{'}h_{1}\right)\sin\left(\gamma_{2}^{'}h_{2}\right)\cos\left(\frac{\pi}{2}\frac{\omega}{\omega_{0}}\right)\right]\nonumber 
\end{align}
\end{widetext}

Although the full expression for the transmission coefficient Eq.~(\ref{eq:T_system-1-1})
is cumbersome, our primary goal is to derive the figures of merit.
First, we determine the transmission magnitude $|T_{\text{system}}|$.
Calculating the magnitude $|T|=\sqrt{\text{Re}^{2}T+\text{Im}^{2}T}$
(up to the second order in $\alpha$, $\beta$ and $\frac{h_{c}}{h}$)
and grouping the terms, we obtain:

\begin{widetext}

\begin{equation}
\begin{aligned}\left|T_{\text{system}}\right|\approx & \cos\left(\frac{\pi}{2}\frac{\omega}{\omega_{0}}+\Phi_{d}\right)+\frac{1}{2}\alpha^{2}\frac{\cos^{2}\left(\frac{\pi}{2}\frac{\omega}{\omega_{0}}\right)\sin^{2}\left(\Phi_{s}\right)}{\cos\left(\frac{\pi}{2}\frac{\omega}{\omega_{0}}+\Phi_{d}\right)}+\\
+ & \alpha^{2}\left[\frac{1}{2}\sin\left(\Phi_{d}\right)\sin\left(\frac{\pi}{2}\frac{\omega}{\omega_{0}}\right)-2\sin\left(\gamma_{1}^{'}h_{1}\right)\sin\left(\gamma_{2}^{'}h_{2}\right)\cos\left(\frac{\pi}{2}\frac{\omega}{\omega_{0}}\right)\right]
\end{aligned}
\end{equation}

\end{widetext}

To simplify the analysis, we express the total residual retardation
phase of the compensated system directly in terms of the mismatch
parameter $\beta$:
\begin{equation}
\frac{\pi}{2}\frac{\omega}{\omega_{0}}+\Phi_{d}\approx\beta\Phi_{s}
\end{equation}
Thus,
\begin{equation}
\begin{split}\left|T_{\text{system}}\right| & \approx\cos\left(\beta\Phi_{s}\right)+\\
+ & \frac{1}{2}\alpha^{2}\frac{\sin^{2}\left(\frac{\pi}{2}\frac{\omega-\omega_{0}}{\omega_{0}}\right)\sin^{2}\left(\Phi_{s}\right)}{\cos\left(\beta\Phi_{s}\right)}+\\
+ & \frac{1}{2}\alpha^{2}\cos\left(\frac{\pi}{2}\frac{\omega-\omega_{0}}{\omega_{0}}-\beta\Phi_{s}\right)\cos\left(\frac{\pi}{2}\frac{\omega-\omega_{0}}{\omega_{0}}\right)+\\
+ & 2\alpha^{2}\sin\left(\gamma_{1}^{'}h_{1}\right)\sin\left(\gamma_{2}^{'}h_{2}\right)\sin\left(\frac{\pi}{2}\frac{\omega-\omega_{0}}{\omega_{0}}\right)
\end{split}
\end{equation}

To estimate the amplitude non-achromaticity ($\mathcal{A}$), defined
as the maximum deviation $1-|T|$, we evaluate this expression under
the worst-case conditions.
\begin{equation}
\mathcal{A}=2\sin^{2}\left(\frac{1}{2}b\Phi_{s}\right)+\frac{1}{2}\alpha^{2}+2\alpha^{2}\left|\sin\left(\frac{\pi}{2}\frac{\omega-\omega_{0}}{\omega_{0}}\right)\right|
\end{equation}
This formula explicitly captures the dependence on the spectral detuning
($\omega-\omega_{0}$) and provides a rigorous upper bound for the
amplitude loss without imposing limits on the magnitude of the arguments.

The phase distortion is dominated by the imaginary term in the transmission
coefficient, which is proportional to $\cos(\Phi_{c})$. Assuming
the QWP retardation follows a linear dispersion law $\Phi_{c}\left(\omega\right)=\frac{\pi}{2}\frac{\omega}{\omega_{0}}$,
this term transforms as: 
\begin{equation}
\cos\left(\Phi_{c}\left(\omega\right)\right)=-\sin\left(\frac{\pi}{2}\frac{\omega-\omega_{0}}{\omega_{0}}\right)
\end{equation}
 Since the distortion is directly proportional to the desired modulation
angle $\varphi$, the Constant Phase Distortion is zero in this approximation:
\begin{equation}
\mathcal{P}_{0}=0
\end{equation}
 For the angular phase distortion $\mathcal{P}_{1}$, we normalize
the phase error by $\varphi$. Considering the worst-case scenario
where the manufacturing tolerance $b$ minimizes the denominator (amplitude
term), we obtain: 
\begin{equation}
\mathcal{P}_{1}=\frac{\left|\sin\left(\frac{\pi}{2}\frac{\omega-\omega_{0}}{\omega_{0}}\right)\right|}{\Phi_{s}\left|\cos\left(b\Phi_{s}\right)\right|}
\end{equation}
This expression demonstrates that the phase fidelity is determined
by the spectral bandwidth (numerator) and the optical thickness (denominator).
It represents the inevitable residual error in a passive system where
the dispersion of the compensating element is utilized to balance
the amplitude response, making perfect cancellation impossible over
a finite frequency range.

\section{\label{subsec:Derivation-of-Figures}Derivation of Figures of Merit
for Strategy C}

To derive the performance metrics for the narrow-band optimization
strategy, we start from the rigorous second-order approximation of
the Jones matrix element $T_{11}$, derived in Appendix \ref{subsec:Supplementary_2}
Eq.~(\ref{eq:T11}). The transmission coefficient is:
\begin{equation}
\begin{aligned}T_{11}= & \exp\left(i\varphi\right)\left[\cos\left(\Phi_{d}\right)-2\alpha^{2}\sin\left(\gamma_{1}^{'}h_{1}\right)\sin\left(\gamma_{2}^{'}h_{2}\right)\right]-\\
- & i\exp\left(i\varphi\right)\left[\alpha\sin\left(\Phi_{s}\right)+\alpha\beta\sin\left(\Phi_{d}\right)\right]
\end{aligned}
\end{equation}

We represent this complex expression in the polar form $T_{\text{CTN}}=|T_{\text{CTN}}|\exp\left(i\varphi\right)\exp\left(i\delta\psi\right)$,
separating the amplitude and the phase distortion components. Assuming
the phase perturbation $\delta\psi$ is small, we can approximate
$\delta\psi\approx\tan\left(\delta\psi\right)=\text{Im}\left(T_{11}e^{-i\varphi}\right)/\text{Re}\left(T_{11}e^{-i\varphi}\right)$.
This yields:

\begin{equation}
\begin{gathered}T_{\text{CTN}}\approx|T_{\text{CTN}}|\exp\left(i\varphi\right)\times\\
\times\exp\left(-i\frac{\alpha\sin\left(\Phi_{s}\right)+\alpha\beta\sin\left(\Phi_{d}\right)}{\cos\left(\Phi_{d}\right)-2\alpha^{2}\sin\left(\gamma_{1}^{'}h_{1}\right)\sin\left(\gamma_{2}^{'}h_{2}\right)}\right)
\end{gathered}
\end{equation}

The amplitude magnitude $|T_{\text{CTN}}|$, calculated as the square
root of the intensity up to the second order in small parameters,
is given by: 
\begin{equation}
\begin{aligned}|T_{\text{CTN}}|\approx & \cos\left(\Phi_{d}\right)+\frac{1}{2}\alpha^{2}\sin^{2}\left(\Phi_{s}\right)-\\
- & 2\alpha^{2}\sin\left(\gamma_{1}^{'}h_{1}\right)\sin\left(\gamma_{2}^{'}h_{2}\right)
\end{aligned}
\end{equation}

For the analysis of Strategy C, we consider the regime of small retardation
mismatch ($\Phi_{d}\ll1$), allowing us to expand $\cos\left(\Phi_{d}\right)\approx1-\Phi_{d}^{2}/2$
and approximate the geometric sine terms as $\sin\left(\gamma_{1}'h_{1}\right)\sin\left(\gamma_{2}'h_{2}\right)\approx\sin^{2}\left(\Phi_{s}/2\right)$.
Under these assumptions, the expressions simplify significantly:

\begin{equation}
|T_{\text{CTN}}|\approx1-\frac{1}{2}\Phi_{d}^{2}-2\alpha^{2}\sin^{2}\left(\frac{\Phi_{s}}{2}\right)+\frac{1}{2}\alpha^{2}\sin^{2}\left(\Phi_{s}\right)\label{eq:Tctn_sm4}
\end{equation}

For the phase term, retaining only the dominant contributions up to
the second order (neglecting terms like $\alpha\beta\Phi_{d}$ which
are third-order), the expression reduces to a cleaner form dependent
on the cumulative retardation $\Phi_{s}$:

\begin{equation}
T_{\text{CTN}}\approx|T_{\text{CTN}}|\exp\left(i\varphi\left[1-\frac{\sin\left(\Phi_{s}\right)}{\Phi_{s}}\right]\right)
\end{equation}
 Here we used the relation $\alpha=\varphi\Phi_{s}^{-1}\left(1-\beta^{2}\right)^{-1}$.
For narrow bandwidth applications, we assume $\frac{\Delta\omega}{\omega_{0}}\Phi_{s}\ll1$.

\paragraph{Optimization of the Operating Point}

We introduce the effective phase modulation $\varphi_{\text{mod}}$
distinct from the director twist $\varphi$: 
\begin{equation}
\varphi_{\text{mod}}=\varphi\left[1-\frac{\sin\left(\Phi_{s0}\right)}{\Phi_{s0}}\right]\label{eq:phi_mod_def}
\end{equation}
 where $\Phi_{s0}=\Phi_{s}\left(\omega_{0}\right)$. To minimize chromaticity,
we select $\Phi_{s0}$ to be an extremum of the function $f\left(x\right)=\sin\left(x\right)/x$.
Setting the derivative to zero yields the condition 
\begin{equation}
\tan\left(\Phi_{s0}^{\text{opt}}\right)=\Phi_{s0}^{\text{opt}}\label{eq:trans}
\end{equation}
To ensure the modulation scaling factor in Eq.~(\ref{eq:phi_mod_def})
is greater than unity (maximizing the achievable phase stroke), we
specifically target the local minima of the sinc function. We approximate
the solution as: 
\begin{equation}
\Phi_{s0}^{\text{opt}}=2\pi n-\frac{\pi}{2}+\delta\label{eq:Fs_opt}
\end{equation}
 To find the value of $\delta$, we can employ the fixed-point iteration
method. Rearranging the transcendental Eq.~(\ref{eq:trans}), we
define the iteration function as: 
\begin{equation}
\delta=-\arcsin\left[\frac{\cos\delta}{2\pi n-\frac{\pi}{2}+\delta}\right]
\end{equation}
 The first-order approximation yields: $\delta_{1}\approx-\left(2\pi n-\pi/2\right)^{-1}$.

However, in a real device, the actual retardation will deviate from
this optimal target due to manufacturing inaccuracy. We define the
actual retardation as: 
\begin{equation}
\Phi_{s0}=\Phi_{s0}^{\text{opt}}+\Delta\Phi\label{eq:Opt2}
\end{equation}
 where the bracketed term is the theoretical optimum, and $\Delta\Phi$
represents the deviation due to thickness inaccuracy. In the worst-case
scenario, $\Delta\Phi=\pm b\Phi_{s0}$.

\paragraph{Derivation of Angular Phase Distortion ($\mathcal{P}_{1}$)}

The chromatic perturbation of the modulation phase, $\delta\varphi_{\text{mod}}\left(\omega\right)$,
is defined as the deviation of the actual phase from the target $\varphi_{\text{mod}}$
across the bandwidth: 
\begin{equation}
\delta\varphi_{\text{mod}}=\varphi_{\text{mod}}\left[\frac{1-\text{sinc}\left(\Phi_{s}\left(\omega\right)\right)}{1-\text{sinc}\left(\Phi_{s0}\right)}-1\right]
\end{equation}
 where $\text{sinc}\left(x\right)=\sin\left(x\right)/x$. Substituting
$\Phi_{s}\left(\omega\right)\approx\Phi_{s0}\left(1+\frac{\omega-\omega_{0}}{\omega_{0}}\right)$
and expanding the expression in powers of the small parameter $\epsilon=\frac{\omega-\omega_{0}}{\omega_{0}}$,
we perform a Taylor series expansion around the optimized point defined
in Eq.~(\ref{eq:Opt2}).

Crucially, because we are operating near a local extremum, the first-order
term proportional to $\epsilon$ vanishes for the ideal thickness.
The remaining linear dependence arises solely from the thickness error
$\Delta\Phi$. The dominant higher-order term is quadratic ($\epsilon^{2}$).
After algebraic simplification and neglecting negligible terms (assuming
$\Delta\Phi\ll1$ ), the phase distortion normalized by the target
phase becomes: 
\begin{equation}
\frac{\delta\varphi_{\text{mod}}}{\varphi_{\text{mod}}}\approx\frac{1}{2}\Phi_{s0}^{\text{opt}}\left(\frac{\omega-\omega_{0}}{\omega_{0}}\right)^{2}-\Delta\Phi\left(\frac{\omega-\omega_{0}}{\omega_{0}}\right)
\end{equation}
 Defining $\mathcal{P}_{1}$ as the maximum deviation over the full
bandwidth $\Delta\omega$, we arrive at: 
\begin{equation}
\mathcal{P}_{1}=\frac{1}{8}\Phi_{s0}^{\text{opt}}\left(\frac{\Delta\omega}{\omega_{0}}\right)^{2}+\frac{1}{2}b\Phi_{s0}^{\text{opt}}\frac{\Delta\omega}{\omega_{0}}
\end{equation}

\paragraph{Derivation of Constant Phase Distortion ($\mathcal{P}_{0}$)}

In Strategy C, the cell thickness is optimized to satisfy the extremum
condition $\Phi_{s0}=\Phi_{s0}^{\text{opt}}$ at a central frequency
$\omega_{0}$. However, the cumulative retardation $\Phi_{s0}$ depends
not only on the cell thickness $h$ and anisotropy $\gamma$, but
also on the director twist angle $\varphi$ (via the twist wave number
$q=\varphi\left(2h\right)^{-1}$). Explicitly: 
\begin{equation}
\Phi_{s0}\left(\varphi\right)=2h\sqrt{q^{2}+\gamma^{2}}=\sqrt{\varphi^{2}+4h^{2}\gamma^{2}}
\end{equation}
 Since the physical thickness $h$ is uniform, the optimization condition
$\Phi_{s0}\left(\varphi\right)=\Phi_{s0}^{\text{opt}}$ can only be
strictly satisfied for a specific design angle $\varphi^{\text{opt}}$.
For all other angles, a deviation $\delta\Phi\left(\varphi\right)$
arises. Assuming the Mauguin regime ($2h\gamma\gg\varphi$), the first-order
Taylor expansion yields: 
\begin{equation}
\delta\Phi\left(\varphi\right)\overset{\text{def}}{=}\Phi_{s0}\left(\varphi\right)-\Phi_{s0}\left(\varphi^{\text{opt}}\right)\approx\frac{\varphi^{2}-\left(\varphi^{\text{opt}}\right)^{2}}{2\Phi_{s0}^{\text{opt}}}\label{eq:dFi}
\end{equation}

This deviation leads to a residual phase error. We define the Constant
Phase Distortion parameter $\mathcal{P}_{0}$ based on the maximum
contribution of this effect over the spectral bandwidth $\Delta\omega$:
\begin{equation}
\mathcal{P}_{0}=\frac{1}{2}\frac{\Delta\omega}{\omega_{0}}\max_{\varphi}\left(|\varphi|\cdot|\delta\Phi\left(\varphi\right)|\right)\label{eq:P0do}
\end{equation}
 Here we approximated $\varphi_{\text{mod}}\approx\varphi$. Substituting
Eq.~(\ref{eq:dFi}) into the definition of $\mathcal{P}_{0}$ Eq.~(\ref{eq:P0do})
we have
\begin{equation}
\mathcal{P}_{0}=\frac{1}{4\Phi_{s0}^{\text{opt}}}\frac{\Delta\omega}{\omega_{0}}\max_{\varphi\in[0,\pi]}\left|\varphi\left(\varphi^{2}-\left(\varphi^{\text{opt}}\right)^{2}\right)\right|\label{eq:P0_opt}
\end{equation}
To minimize the Constant Phase Distortion $\mathcal{P}_{0}$, we must
select the optimal design angle $\varphi^{\text{opt}}$ that minimizes
the maximum value of function $f=\varphi\left(\varphi^{2}-\left(\varphi^{\text{opt}}\right)^{2}\right)$
over the full modulation range $\varphi\in\left[0,\pi\right]$. This
is a minimax problem. The maximum error occurs either at the local
extremum of the function ($\varphi_{\text{peak}}$) or at the boundary
of the domain ($\varphi=\pi$). The derivative of $f$ vanishes at
$\varphi_{\text{peak}}=\varphi^{\text{opt}}/\sqrt{3}$. The error
at this point is: 
\begin{equation}
\left|f\left(\varphi_{\text{peak}}\right)\right|=\frac{2}{3\sqrt{3}}\left(\varphi^{\text{opt}}\right)^{3}
\end{equation}
the error at the maximum twist angle $\varphi=\pi$ is: 
\begin{equation}
\left|f\left(\pi\right)\right|=\pi\left|\pi^{2}-\left(\varphi^{\text{opt}}\right)^{2}\right|
\end{equation}
The optimal solution is found by equating the error magnitudes at
the peak and the boundary: $\left|f\left(\varphi_{\text{peak}}\right)\right|=\left|f\left(\pi\right)\right|$.
Solving this equation yields the analytical solution: 
\begin{equation}
\varphi^{\text{opt}}=\frac{\sqrt{3}}{2}\pi\label{eq:Fiopt}
\end{equation}
To retrieve the $\mathcal{P}_{0}$ we should substitute $\varphi^{\text{opt}}$
Eq.~(\ref{eq:Fiopt}) and $\varphi=\pi$ back into the expression
Eq.~(\ref{eq:P0_opt}):

\begin{equation}
\mathcal{P}_{0}=\frac{\pi^{3}}{16\Phi_{s0}}\frac{\Delta\omega}{\omega_{0}}
\end{equation}

\paragraph{Derivation of Amplitude Non-Achromaticity ($\mathcal{A}$)}

To estimate the amplitude fidelity, we evaluate the magnitude expression
Eq.~(\ref{eq:Tctn_sm4}) at the optimized operating point $\Phi_{s0}^{\text{opt}}\approx2\pi n-\frac{\pi}{2}$
Eq.~(\ref{eq:Fs_opt}). Substituting the worst-case retardation mismatch
$\Phi_{d}\approx b\Phi_{s0}$, we obtain the final estimate for the
amplitude non-achromaticity
\begin{equation}
\mathcal{A}=\frac{1}{2}\left(b\Phi_{s0}\right)^{2}+\frac{1}{2}\alpha^{2}
\end{equation}

This confirms that the amplitude loss is driven quadratically by both
the manufacturing tolerance and the fundamental non-adiabaticity.
\end{document}